\documentclass[twocolumn]{aastex62}
\usepackage{natbib}
\usepackage{graphicx}
\usepackage{graphics}
\usepackage{amssymb}
\usepackage{amsmath}
\usepackage{mathtools}
\usepackage{multirow}
\usepackage{bm}
\usepackage{braket}
\usepackage{nicefrac}

\usepackage{mathrsfs}

\usepackage[utf8]{inputenc}
\usepackage[T1]{fontenc}

\def\utw{\smash{\rlap{\lower5pt\hbox{$\sim$}}}}
\def\udtw{\smash{\rlap{\lower6pt\hbox{$\approx$}}}}

\shorttitle{QPOs in Swift/BAT GRBs}
\shortauthors{Tarnopolski \& Marchenko}

\newcommand{\lesssimgtr}{\mathrel{\mathpalette\lagA{<>}}}

\newcommand{\lagA}[2]{\lagB#1#2}
\newcommand{\lagB}[3]{%
  \vcenter{\offinterlineskip\mathsurround=0pt
    \halign{##\cr$#1#2$\cr\noalign{\vskip0.2pt}$#1\sim$\cr\noalign{\vskip-1.5pt}$#1#3$\cr}%
  }%
}

\DeclarePairedDelimiter\floor{\lfloor}{\rfloor}
\DeclarePairedDelimiter\ceil{\lceil}{\rceil}

\begin{document}

\title{A Comprehensive Power Spectral Density Analysis of Astronomical Time Series. II. The Swift/BAT Long Gamma-Ray Bursts}

\author[0000-0003-4666-0154]{Mariusz Tarnopolski}
\email{mariusz.tarnopolski@uj.edu.pl}
\affiliation{Astronomical Observatory, Jagiellonian University, Orla 171, 30--244, Krak\'ow, Poland}

\author[0000-0002-7175-1923]{Volodymyr Marchenko}
\email{volodymyr.marchenko@oa.uj.edu.pl}
\affiliation{Astronomical Observatory, Jagiellonian University, Orla 171, 30--244, Krak\'ow, Poland}

\begin{abstract}

We investigated the prompt light curves (LCs) of long gamma-ray bursts (GRBs) from the Swift/BAT catalog. We aimed to characterize their power spectral densities (PSDs), search for quasiperiodic oscillations (QPOs), and conduct novel analyses directly in the time domain.
We analyzed the PSDs using Lomb-Scargle periodograms, and searched for QPOs using wavelet scalograms. We also attempted to classify the GRBs using the Hurst exponent, $H$, and the $\mathcal{A}-\mathcal{T}$ plane.
The PSDs fall into three categories: power law (PL; $P(f)\propto 1/f^\beta$) with index $\beta\in(0,2)$, PL with a non-negligible Poisson noise level (PLC) with $\beta\in(1,3)$, and a smoothly broken PL (SBPL; with Poisson noise level) yielding high-frequency index $\beta_2\in(2,6)$. The latter yields break time scales on the order of 1--100\,seconds. The PL and PLC models are broadly consistent with a fully developed turbulence, $\beta=\nicefrac{5}{3}$. For an overwhelming majority of GRBs (93\%), $H>0.5$, implying ubiquity of the long-term memory. We find no convincing substructure in the $\mathcal{A}-\mathcal{T}$ plane. Finally, we report on 34 new QPOs: with one or more constant leading periods, as well as several chirping signals. The presence of breaks and QPOs suggests the existence of characteristic time scales that in at least some GRBs might be related to the dynamical properties of plasma trajectories in the accretion disks powering the relativistic jets.

\end{abstract}


\section{Introduction}
\label{sect1}

Gamma-ray bursts \citep[GRBs,][]{klebesadel} are typically divided into {\it short}, coming from double neutron star (NS-NS) or NS-black hole (BH) mergers \citep{nakar07,berger11}, and {\it long} ones, whose progenitors are the collapse of massive stars, e.g. Wolf-Rayet or blue supergiants \citep{woosley06,cano17}. The division between the two classes is primarily based on the bimodal distribution of $T_{90}$ \citep[time during which 90\% of the GRB's fluence is accumulated;][]{kouveliotou93}, and the threshold is at $T_{90}\simeq 2\,{\rm s}$ (but cf. \citealt{bromberg13,tarnopolski15a}). GRBs exhibit a rich variety of light curve (LC) shapes \citep{fishman94}, which implies complicated mechanisms governing their radiation. The LCs usually exhibit a power law (PL; $P(f)\propto 1/f^\beta$) power spectral density (PSD), with or without a break, and on rare occasions a sign of a quasiperiodic oscillation (QPO) was noted.

The first confirmed QPO in a GRB was found in the 5 March 1979 event \citep{barat79,terrell80}, in which an unambiguous $\sim$8~s periodicity (lasting for $\sim$20 cycles) followed a 0.2~s outburst. However, subsequent analyses \citep{norris91,fenimore96} suggested that this transient was actually a soft gamma repeater (SGR). Shortly later, a 4.2~s periodicity was reported in a 29 October 1977 event during about 30~s of its duration \citep{wood81}, while \citet{kouveliotou88} identified 7 cycles of a 2.2~s quasiperiodicity in a long (43~s), hard (extending to 100\,MeV) GRB observed on 5 August 1984. \citet{schaefer88} examined the significance of periods (in the range 2--18~s) claimed for 16 GRBs at that time. They confirmed only the $\sim$8~s periodicity in the 5 March 1979 event. Subsequently, a period of 2.2~s was identified in an 11 May 1988 event \citep{owens90}.

GRB 090709A gained attention when a QPO with a period of $\sim$8~s was reported based on Swift, Konus, Suzaku, and INTEGRAL observations \citep{markwardt09,golenetskii09,gotz09,ohno09}. However, subsequent analyses of the Swift and INTEGRAL \citep{deluca10}, and Swift and Suzaku data \citep{cenko10} did not confirm it on a 3$\sigma$ significance level, but another reanalysis of the Suzaku LC revealed a 3$\sigma$-significant QPO \citep{iwakiri10}. Likewise it was found that the Swift LC actually exhibits a 3.5$\sigma$ significance for the QPO \citep{ziaeepour11}. Again, like in the 5 March 1979 event, such a periodicity might hint at an SGR nature of the source, however in this case all other properties ensure its GRB origin. The presumed QPO was speculated to be caused by magnetorotational instabilities (MRI) in a hyperaccreting disk \citep{masada07}, or originate due to a precessing magnetic field \citep{ziaeepour11}. While an unambiguous conclusion about this QPO has apparently not been reached, the possibility of such modulations in the prompt emission of GRBs is fascinating. Moreover, \citet{beskin10} discovered the first periodic pulsations in the optical prompt emission of GRB 080319B, at a period of 8.1~s. The presumed (quasi)periodic nature ought to be taken with caution, though, since the detected period corresponds to only four peaks in the LC.

First searches for high-frequency QPOs were unsuccesful \citep{deng97,kruger02}, but \citet{zhilyaev09} employed a wavelet-based approach to short BATSE GRBs, which yielded a QPO with a leading period of 5.5~ms in one case (trigger number 207), and a few chirping signals as well. NS-BH mergers are indeed expected to give rise to jet precession triggering the QPO modulation \citep{stone13}. However, a subsequent canonical search for QPO features in PSDs detected no significant signals \citep{dichiara13b}.

When it comes to the overall shape of the PSD, \citet{belli92} observed PL ($1\lesssim\beta\lesssim 2$) or Lorentzian (i.e., indicative of an autoregressive process of order 1, i.e. AR(1)) forms in case of 5 long Konus GRBs. \citet{giblin98} examined 100 GRBs by computing their PSDs and fitting a PL. They obtained a wide range of $\beta$, extending up to $\beta\approx 7$, with 65\% of the cases exceeding the red noise value, i.e. $\beta>2$. 

\citet{beloborodov98}, in turn, constructed the average PSD of 214 certainly long GRBs ($T_{90}>20\,{\rm s}$), hence considered the LCs as random realizations of the same underlying stochastic process, and concluded that the obtained $\beta\simeq\nicefrac{5}{3}$ is consistent with a fully developed turbulence\footnote{Through Parseval's theorem, the energy at frequency $f$ can be expressed as the Fourier transform of the signal, and the wavenumber $k\propto f$. Hence it follows that for turbulence the PSD has an exponent of $\nicefrac{5}{3}$ \citep[cf.][]{moraghan15}.}, arising in the internal shock model that likely governs the observed variability. A subsequent analysis with a bigger sample of 514 GRBs confirmed the Kolmogorov's $\nicefrac{5}{3}$ law for the average PSD \citep{beloborodov2000}, and showed that dim bursts exhibit steeper PSDs \citep[cf.][]{ryde03,rong03}. Also, in lower energy chanels the PSDs are steeper than at higher energies. \citet{panaitescu99}, on the other hand, modeled the PSD based on the internal shock model, and found that the $\nicefrac{5}{3}$ law can be explained by invoking modulation of the relativistic winds instead of turbulence. \citet{pozanenko2000} computed the average PSD of 815 long GRBs and fitted it with a PL and an exponential PL with $\beta\in(\nicefrac{4}{3},\nicefrac{5}{3})$, again roughly indicative of the Kolmogorov law. \citet{chang2000} simulated GRB LCs as a sum of fast-rise-exponential-decay (FRED) pulses, and demonstrated that by adjusting the sampling, rise and decay time scales, the $\nicefrac{5}{3}$ law can be recovered. With a Swift sample of GRBs with redshifts, \citet{guidorzi12} found that the $\nicefrac{5}{3}$ law holds in the rest frame as well (roughly, as $\beta\lesssim 2$ depending on the subsample---for higher redshifts the PSD becomes shallower), and identified a break in the smoothly broken PL at time scales $\sim$30~s. \citet{dichiara13a} arrived at similar results for GRBs observed with BeppoSAX and Fermi, with a break at $\sim 15-25$~s, depending on the energy channel. The consistency and persistence of the average PSD index of $\nicefrac{5}{3}$ among different data sets, corresponding to different energy bands, instrument sensitivities etc., strongly suggest that indeed the collection of GRBs shall be treated as an ensemble, and that each LC is a stochastic realization of the same underlying emission mechanism (though possibly generated by values of parameters different from burst to burst, owing to, e.g., different magnetization degrees). \citet{guidorzi16} analyzed the individual PSDs of Swift GRBs, finding that they can be divided into two classes: with and without a break. The overall span of the PL index fell into $1.2\lesssim\beta\lesssim 4$, while the break time scales spanned the range $\sim 1-180$~s, with a logarithmic average of 25~s. They also found marginal evidence (barely touching the 3$\sigma$ level) for QPOs in three cases. Finally, \citet{dichiara16} found a statistically significant anticorrelation between the rest-frame peak energy, $E_{\rm peak}^{\rm rest}$, and the PSD index $\beta$, adding to the long list of correlations between various parameters of the prompt and afterglow phases \citep{shahmoradi15,dainotti17,dainotti18a,dainotti18b}. The $E_{\rm peak}^{\rm rest}-\beta$ relation was discussed on grounds of a few prompt emission models, explaining the overall anticorrelation by invoking the bulk Lorentz factor, $\Gamma$, as the key observable connecting the two quantities. The distribution of the break time scales spanned the range $\sim 1-140$~s, with a logarithmic mean of 20~s. \citet{boci18} suggested that the $E_{\rm peak}^{\rm rest}-\beta$ relation might be a consequence of the peak energy--luminosity and duration--luminosity relations.

\citet{zhang14} simulated several LCs within the Internal-Collision-induced MAgnetic Reconnection and Turbulence (ICMART) model, with moderately high magnetization of the ejected shells. It was found that for various reasonable parameter values, the resulting PSDs were generally consistent with the PL forms of Swift GRBs. In particular, $1\lesssim\beta\lesssim 2$ for the assumed range of ratio of the mini-jets and $\Gamma$ factors (although for some particular parameters, occasionally steeper PSDs were also arrived at). Additionally, spikier LCs (more variable on shorter time scales) yielded shallower PSDs.

The Hurst exponent, $H$ (a measure of persistence, or self-similarity of a time series), was shown to be able to differentiate between short and long GRBs  \citep{maclachlan13}, especially when coupled with other characteristics, like duration $T_{90}$ and minimum variability time scale \citep[MVTS;][]{tarnopolski15c}. $H$ is constrained to the interval $(0,1)$, and is applicable to both stationary and nonstationary time series, hence is able to provide a universal classification of GRBs. Such a classification, based predominantly on their LCs, was notoriously difficult due to the diverse morphology. Recently, \citet{jespersen20} applied a machine learning dimensionality reduction algorithm, t-distributed stochastic neighborhood embedding (t-SNE) to Swift GRBs. t-SNE groups similar LCs close together, based on which it was demonstrated that as a result two prominent clusters emerged, each corresponding to the short and long subclasses, respectively. This appears to resolve the issue whether there are two or three main GRB types \citep{horvath02,horvath08,tarnopolski16a,horvath19,toth19,tarnopolski19a,tarnopolski19b,tarnopolski19c}. 

The goal of this paper is a possibly comprehensive analysis of PSDs of a big sample of GRBs from the Swift catalog, with a particular aim to identify QPOs. In addition, the LCs are investigated directly in the time domain with the Hurst exponent, and the recently developed $\mathcal{A}-\mathcal{T}$ plane, which was proven to be capable to classify blazar subtypes based solely on the LCs \citep{tarnopolski20b}. In Sect.~\ref{sect2} the description of the utilized GRB sample is given, and the outline of the employed analysis methods is provided. In Sect.~\ref{sect3} the results are presented, and Sect.~\ref{sect4} is devoted to discussion. Concluding remarks are gathered in Sect.~\ref{sect5}.

\section{Data and methods}
\label{sect2}

Description of the sample is given first. The employed methods are then briefly described. For a more detailed explanation, as well as the results of a comprehensive benchmark testing of each method, we refer the reader to \citet{tarnopolski20b}.

\subsection{Sample}
\label{sect2.1}

The mask-weighted, background-subtracted LCs, in a 64-ms binning and covering the total energy range $15-350$~keV, were downloaded from the Swift/BAT catalog\footnote{\url{https://swift.gsfc.nasa.gov/results/batgrbcat/}} \citep{lien16}. The portions of the LCs within respetive $T_{100}$ intervals were extracted. We focus on long GRBs with sufficient number of points to conduct a meaningful time series and PSD analysis. We therefore utilized LCs with more than 50 points\footnote{This is a requirement of the software \textsc{wavepal} used for the wavelet scalograms; cf. Sect.~\ref{sect2.3}.}, i.e. with $T_{100} > 3.2\,{\rm s}$. We excluded confirmed short GRBs with extended emission.\footnote{\url{https://swift.gsfc.nasa.gov/results/batgrbcat/summary_cflux/summary_GRBlist/GRBlist_short_GRB_with_EE.txt}} We ended with 1160 GRBs in our sample.

\subsection{PSDs}
\label{sect2.2}

\subsubsection{Lomb-Scargle periodogram}

To calculate the PSD of a time series $\{x_k(t_k)\}_{k=1}^n$ with a constant time interval between consecutive observations, $\delta t = t_{k+1} - t_k \equiv 64\,{\rm ms}$, the Lomb-Scargle periodogram \citep[LSP; ][]{lomb,scargle,vanderplas18} is computed in the standard way as
\begin{equation}
\begin{split}
P_{LS}(\omega) = \frac{1}{2\sigma^2} &\left[ \frac{\left(\sum\limits_{k=1}^n (x_k-\bar{x}) \cos[\omega(t_k-\mathscr{T})]\right)^2}{\sum\limits_{k=1}^n \cos^2[\omega(t_k-\mathscr{T})]} \right. \\
+& \left. \frac{\left(\sum\limits_{k=1}^n (x_k-\bar{x}) \sin[\omega(t_k-\mathscr{T})]\right)^2}{\sum\limits_{k=1}^n \sin^2[\omega(t_k-\mathscr{T})]}  \right],
\end{split}
\label{eq1}
\end{equation}
where $\omega = 2\pi f$ is the angular frequency, $\mathscr{T}\equiv\mathscr{T}(\omega)$ is
\begin{equation}
\mathscr{T}(\omega)=\frac{1}{2\omega} \arctan \left[ \frac{\sum\limits_{k=1}^n\sin(2\omega t_k)}{\sum\limits_{k=1}^n\cos(2\omega t_k)} \right],
\label{eq2}
\end{equation}
and $\bar{x}$ and $\sigma^2$ are the sample mean and variance.

The lower limit for the sampled frequencies is $f_{\rm min} = 1/(t_{\rm max}-t_{\rm min})$, corresponding to the length of the time series. Since we are dealing with uniformly sampled data, the upper limit is the Nyquist frequency, $f_{\rm max}\equiv f_{\rm Nyq} = \frac{1}{2\delta t}$. The total number of sampled frequencies is
\begin{equation}
N_P = n_0\frac{f_{\rm max}}{f_{\rm min}},
\label{eq3}
\end{equation}
and we employ $n_0 = 100$ hereinafter.

\subsubsection{Binning and fitting}

For fitting in the log-log space, binning is applied. The values of $\log f$ are binned into approximately equal-width bins, with at least two points in a bin, and the representative frequencies are computed as the geometric mean in each bin. The PSD value in a bin is taken as the arithmetic mean of the logarithms of the PSD \citep{papadakis93,isobe15}. We require the binned PSDs to consist of at least seven points for the fitting, which left us with 1150 GRBs suitable for the PSD analysis.

The following models were fitted to the binned LSPs:
\begin{enumerate}
\item pure power law (PL):
\begin{equation}
P(f) = P_{\rm norm}f^{-\beta},
\label{eq4}
\end{equation}
\item PL plus Poisson noise (PLC):
\begin{equation}
P(f) = P_{\rm norm}f^{-\beta} + C,
\label{eq5}
\end{equation}
\item smoothly broken PL \citep[SBPL;][]{mchardy2004} plus Poisson noise:
\begin{equation}
P(f) = \frac{P_{\rm norm}f^{-\beta_1}}{1 + \left( \frac{f}{f_{\rm break}}\right)^{\beta_2-\beta_1}} + C,
\label{eq6}
\end{equation}
\item SBPL plus Poisson noise, with a fixed $\beta_1=0$:\footnote{When $\beta_2=2$, this is a Lorentzian (plus Poisson noise), i.e. a PSD of an AR(1) process. }
\begin{equation}
P(f) = \frac{P_{\rm norm}}{1 + \left( \frac{f}{f_{\rm break}}\right)^{\beta_2}} + C,
\label{eq6a}
\end{equation}
\end{enumerate}
where the parameter $C$ is an estimate of the Poisson noise level coming from the uncertainties of individual measurements (see further in this Section), $\beta$ is the PL index, $f_{\rm break}$ is the break frequency (from which the break time scale is calculated as $T_{\rm break}=1/f_{\rm break}$), and $\beta_1,\,\beta_2$ are the low- and high-frequency indices, respectively. PL is a case of PLC with $C=0$. The reason for considering them separately is that the PLC model degenerates when $\beta_{PLC}\rightarrow 0$, since $P(f)\rightarrow P_{\rm norm}+C=\rm{const.}$ then, and hence fitting a pure PL diminishes the parameter uncertainties (cf. \citealt{zywucka20}). Similarly, SBPL was found to often take advantage of the degree of freedom provided by the possibility to vary $\beta_1$, and led to overfitting, hence the two variants of the SBPL were considered separately as well. For completeness, an SBPL with $\beta_1=\beta_2$ reduces to a PLC.

Fits of different models were compared using the small sample Akaike information criterion ($AIC_c$) given by
\begin{equation}
AIC_c=2p-2\mathcal{L}+\frac{2(p+1)(p+2)}{N-p-2},
\label{}
\end{equation}
where $\mathcal{L}$ is the log-likelihood, $p$ is the number of parameters, and $N$ is the number of points fitted to \citep{akaike74,hurvich89,burnham04}. For a regression problem,
\begin{equation}
\mathcal{L} = -\frac{1}{2}N\ln\frac{{\rm RSS}}{N},
\label{}
\end{equation}
where RSS is the residual sum of squares; $p$ is an implicit variable in $\mathcal{L}$. A preferred model is one that minimizes $AIC_c$. This criterion is a trade-off between the goodness of fit and the complexity of the model, expressed via the number of parameters $p$. What is essential in assesing the relative goodness of a fit in the $AIC_c$ method is the difference, $\Delta_i=AIC_{c,i}-AIC_{c,\rm min}$, between the $AIC_c$ of the $i$th model and the one with the minimal $AIC_c$. If $\Delta_i<2$, then there is substantial support for the $i$th model (or the evidence against it is worth only a bare mention), and the proposition that it is a proper description is highly probable. If $2<\Delta_i<4$, then there is strong support for the $i$th model. When $4<\Delta_i<7$, there is considerably less support, and models with $\Delta_i>10$ have essentially no support.

The MVTS, $\tau$, is defined herein as the time scale (corresponding to a frequency $f_0 = 1/\tau$) at which the Poisson noise level dominates over the PL/SBPL component. E.g., for the PLC case it is obtained by solving the equation $P_{\rm norm}f_0^{-\beta} = C$, and similarly for the SBPL case (which, however, does not yield a closed-form solution, hence is obtained numerically). The standard errors of $f_0$ are estimated via bootstrapping: 1000 random realizations of the best-fit PSD were generated by varying the parameters within their uncertainties, and the standard deviation of the resulting sample was calculated. We record only cases with $\Delta\tau<\tau$.

The Poisson noise level, coming from the statistical noise due to uncertainties in the LC's observations, $\Delta x_k$, is the mean squared error, with a normalization suitable for LSP:
\begin{equation}
P_{\rm Poisson} = \frac{1}{2\sigma^2 n} \sum\limits_{k=1}^n \Delta x_k^2.
\label{eq6b}
\end{equation}

\subsection{QPOs}
\label{sect2.3}

To search for QPOs, we employ the wavelet scalogram. A wavelet $\psi(t)$ is a wave packet, i.e. its location and instantaneous frequency are well constrained. We use the Morlet wavelet hereinafter,
\begin{equation}
\psi(t) = \frac{1}{\pi^{1/4}} \left[\exp{\left( \textrm{i}\omega_0 t \right)} - \exp{\left( -\frac{\omega_0^2}{2} \right)}\right] \exp{\left(-\frac{t^2}{2}\right)},
\label{eqMorlet}
\end{equation}
with $\omega_0=10-20$ to ensure a good frequency resolution. The mother wavelet gives rise to the dictionary, or child wavelets, $\psi_{s,l}(t)$, into which the analyzed signal is decomposed: $x(t)=\sum_{s,l}W(s,l)\psi_{s,l}(t)$. The coefficients of such a decomposition, $W(s,l)$, depend on the location, $l\in\mathbb{R}$, and the scale, $s\in\mathbb{R}_+$. The scalogram therefore allows not only to identify the frequency, but localize it within the time series in the temporal domain as well, and visualize it in the time-frequency space. For this purpose, we utilize the method implemented in the package \textsc{wavepal}\footnote{\url{https://github.com/guillaumelenoir/WAVEPAL}} \citep{lenoir18a,lenoir18b}. To test the significance of the detected features, they are tested against a continuous-time autoregressive moving average (CARMA) stochastic model \citep{kelly14}. This is a more general family of noise than the easily tested white noise, or commonly considered colored noise. We aim to detect QPOs at the level of at least $3\sigma$ (99.73\% confidence level).

\subsection{Hurst exponents}
\label{sect2.4}

The Hurst exponent $H$ measures the statistical self-similarity of a time series $x(t)$. It is said that $x(t)$ is self-similar (or self-affine) if it satisfies
\begin{equation}
x(t)\stackrel{\textbf{\textrm{.}}}{=}\lambda^{-H}x(\lambda t),
\label{eq7}
\end{equation}
where $\lambda>0$ and $\stackrel{\textbf{\textrm{.}}}{=}$ denotes equality in distribution. The meaning of $H$ can be understood as follows: for a persistent stochastic process, if some measured quantity grows on average (over some time periods), the system prefers to maintain their growth. The process is, however, probabilistic, and hence at some point the observed quantity will eventually start to decrease (on average). But the process still has long-term memory (which is a global feature), therefore it prefers to decrease for some time until the transition occurs randomly again. In other words, the process prefers to sustain its most recent behavior (in a statistical sense). In case of $H<0.5$, the process is anti-persistent, and it possesses short-term memory, meaning that the observed values frequently switch from relatively high to relatively low (with respect to a stationary mean), and there is no preference among the increments. This is a so-called mean-reverting process. A PSD in form of a PL is indicative of a self-similar process. There is a (piecewise) linear relation between $H$ and the index $\beta$ of a PL PSD: $H=(\beta + 1)/2$ for $\beta\in(-1,1)$, and $H=(\beta - 1)/2$ for $\beta\in(1,3)$, with $\beta=0$ (white noise) and $\beta=2$ (red noise) both yielding $H=0.5$. The case $\beta=1$ (pink noise) is at the border, with no precise $H$ value assigned.

We utilize three algorithms\footnote{The \textsc{Mathematica} implementations are available at \url{https://github.com/mariusz-tarnopolski/Hurst-exponent-and-A-T-plane}.} for extracting $H$: the detrended fluctuation analysis (DFA), and two wavelet-based methods: the discrete wavelet transform (DWT) with the Haar wavelet as the basis, and the averaged wavelet coefficient (AWC) method.

\subsubsection{Detrended fluctuation analysis---DFA}
\label{sect2.4.1}

In the DFA algorithm \citep{peng94,peng95} one starts with calculating the accumulative sum
\begin{equation}
X(t) = \sum\limits_{k=1}^t \big( x_k-\bar{x} \big),
\label{eq8}
\end{equation}
which is next partitioned into non-overlapping segments of length $\varsigma$ each. In each segment, the corresponding part of the time series $X(t)$ is replaced with its linear fit, resulting in a piecewise-linear approximation of the whole $X(t)$, denoted by $X_{\rm lin}(t;\varsigma)$. The fluctuation as a function of the segment length $\varsigma$ is defined as
\begin{equation}
F(\varsigma) = \left[\frac{1}{N}\sum\limits_{t=1}^N \left(X(t)-X_{\rm lin}(t;\varsigma)\right)^2\right]^{\nicefrac{1}{2}}.
\label{eq9}
\end{equation}
The slope $a$ of the linear regression of $\log F(\varsigma)$ versus $\log\varsigma$ is an estimate for $H$: $H=a$ if $a\in(0,1)$, and $H=a-1$ if $a\in(1,2)$. 

\subsubsection{Averaged wavelet coefficient---AWC}
\label{sect2.4.2}

The AWC method \citep{simonsen98} relies directly on the scaling in Eq.~(\ref{eq7}) and employs the continuous wavelet transform, which leads to
\begin{equation}
W(\lambda s, \lambda l) = \lambda^{H+1/2} W(s,l).
\label{eq10}
\end{equation}
The AWC is defined as the standard arithmetic mean over the locations $l$ at a given scale $s$:
\begin{equation}
W(s) = \braket{\left| W(s,l) \right|}_l.
\label{eq11}
\end{equation}
By a linear regression of $\log W(s)$ vs. $\log s$, an estimate of $H$ can be obtained from the slope $a$ via $H=a-\nicefrac{1}{2}$ if $a\in(\nicefrac{1}{2},\nicefrac{3}{2})$, and $H=a+\nicefrac{1}{2}$ if $a\in(-\nicefrac{1}{2},\nicefrac{1}{2})$.

\subsubsection{Discrete wavelet transform---DWT}
\label{sect2.4.3}

$H$ can be obtained with the DWT using, e.g., the Haar wavelet as the basis \citep{veitch99,knight17}. The relation between the variance of the wavelet transform coefficients $d_{j,k}$ (where $2^j$ corresponds to the scale $s$, and $k\cdot 2^j$ to the location $l$) and the scale $j$ can be written as
\begin{equation}
\log_2 {\rm var}(d_{j,k})=a\cdot j+{\rm const.}
\label{eq12}
\end{equation}
The slope $a$ is obtained by fitting a line to the linear part of the $\log_2 {\rm var}(d_{j,k})$ vs. $j$ relation, and $H$ is obtained as $H=(a-1)/2$ when $a\in (1,3)$, and $H=(a+1)/2$ when $a\in (-1,1)$.

\subsection{The $\mathcal{A}-\mathcal{T}$ plane}
\label{sect2.5}

The $\mathcal{A-T}$ plane was initially designed to provide a fast and simple estimate of the Hurst exponent \citep{tarnopolski16d}. It is also well suited to differentiate between types of colored noise, Eq.~(\ref{eq4}), characterized by different PL indices $\beta$ \citep{zunino17}, and to discriminate between regular and chaotic dynamics \citep{zhao18}. It comprises of the fraction of turning points, $\mathcal{T}$, and the Abbe value, $\mathcal{A}$.

\subsubsection{Turning points}

Consider three consecutive data points, $x_{k-1},x_k,x_{k+1}$. They can be arranged in six ways; in four of them, they will create a peak or a trough, i.e., a turning point \citep{kendall,brockwell}. The probability of finding a turning point in such a subset is hence $\nicefrac{2}{3}$. Let $\mathcal{T}$ denote the fraction of turning points in a time series comprised of $n$ points. Therefore $\mathcal{T}\in[0,1]$, and is asymptotically equal to $\nicefrac{2}{3}$ for a purely uncorrelated time series (white noise). A process with $\mathcal{T}>\nicefrac{2}{3}$ (i.e., with raggedness exceeding that of a white noise) will be more noisy than white noise. Similarly, a process with $\mathcal{T}<\nicefrac{2}{3}$ will be ragged less than white noise. All of these cases can be realized for various stochastic processes (e.g., PL or autoregressive moving average) as well as real-world instances \citep{bandt07,tarnopolski19d}. 

\subsubsection{Abbe value}

The Abbe value is defined as half the ratio of the mean-square successive difference to the variance \citep{neumann,neumann2,kendall1971}:
\begin{equation}
\mathcal{A} = \frac{\frac{1}{n-1}\sum\limits_{i=1}^{n-1}\left( x_{i+1}-x_i \right)^2}{\frac{2}{n}\sum\limits_{i=1}^n\left( x_i-\bar{x} \right)^2}\equiv \frac{1}{2} \frac{{\rm var}\left( dX \right)}{{\rm var}\left( X \right)},
\label{eq13}
\end{equation}
where $dX$ denotes the increments (consecutive differences) of process $X$. $\mathcal{A}$ quantifies the smoothness (raggedness) of a time series by comparing the sum of the squared differences between two successive measurements (the variance of the differenced process $dX$) with the variance of the whole time series $X$. It approaches zero for time series displaying a high degree of smoothness, while the normalization factor ensures that $\mathcal{A}$ tends to unity for a white noise process \citep{williams}.  In astronomy it has been rarely utilized, with some recent, nonextensive examples (\citealt{Shin09,mowlavi,sokolovsky16,perez17}; but see also \citealt{lafler65}). In particular, it was demonstrated that blazar subclasses, observed in $\gamma$-rays, are separated in the $\mathcal{A}-\mathcal{T}$ plane \citep{tarnopolski20b}, as are optically observed blazar candidates \citep{zywucka20} behind the Magellanic Clouds \citep{zywucka18}.

\begin{figure}
\centering
\includegraphics[width=\columnwidth]{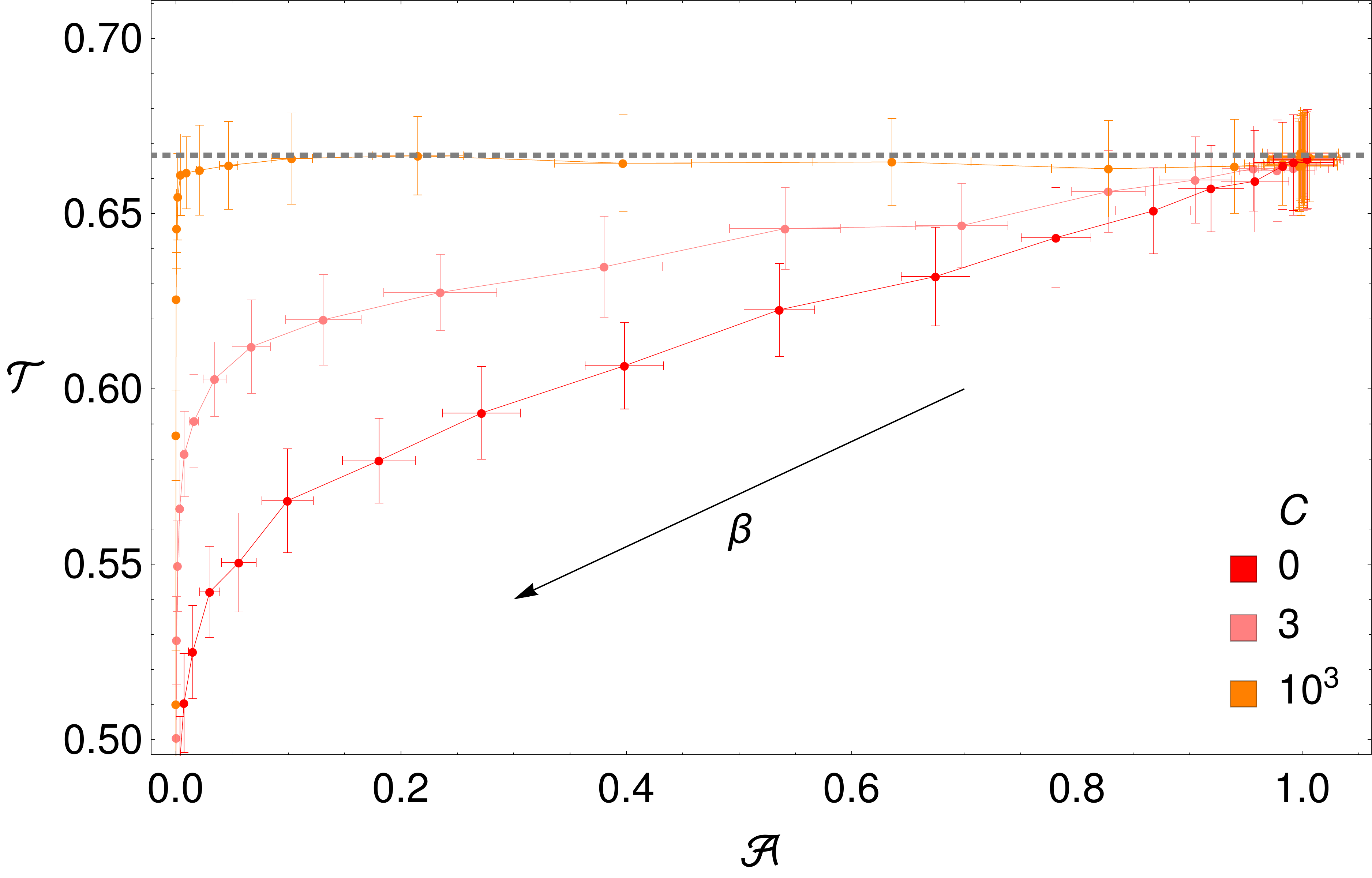}
\caption{Locations in the $\mathcal{A-T}$ plane of the PL plus Poisson noise PSDs of the form $P(f)\propto 1/f^\beta+C$, with $\beta\in\{0,0.1,\ldots,3\}$. For each PSD, 100 realizations of the time series were generated, and the displayed points are the mean locations of them. The error bars depict the standard deviation of $\mathcal{A}$ and $\mathcal{T}$ over these 100 realizations. The case $\beta=0$ is a pure white noise, with $(\mathcal{A},\mathcal{T})=(1,\nicefrac{2}{3})$. The generic PL case ($C=0$) is the lowest curve (red); with an increasing level of the Poisson noise $C$, the curves are raised and shortened, as the white noise component starts to dominate over the PL part.}
\label{fig_AT}
\end{figure}

In Figure~\ref{fig_AT} the locations in the $\mathcal{A}-\mathcal{T}$ plane of PL processes are shown, as well as PLC ones. In effect of introducing Poisson noise $C$, the location $(\mathcal{A},\mathcal{T})$ of the otherwise PL-type signal is being dragged closer to the point $(1,\nicefrac{2}{3})$ corresponding to white noise. Therefore, the region of availability of time series with PSDs of the PLC type is two-dimensional (bounded from below by the PL limit, and from above by the line $\mathcal{T}=\nicefrac{2}{3}$), allowing for nontrivial relations between $\mathcal{A}$ and $\mathcal{T}$.

\subsubsection{Coarse graining}

The so-called coarse-grained sequences are calculated according to \citet{zunino17}. They are obtained by dividing the set $\{x_k\}$ into nonoverlapping segments of length $d$, and each segment is averaged, resulting in smoothed sequences $\{y_j^d\}$:
\begin{equation}
y_j^d = \sum\limits_{k=(j-1)d + 1}^{jd} x_k
\label{eq60}
\end{equation}
for $j\in\{1,\ldots,\lfloor n/d\rfloor \}$. The $\mathcal{A}-\mathcal{T}$ plane, constructed as a function of the temporal scale factor $d$, allows to investigate various temporal resolutions given only one realization of the process, i.e., an LC in only one binning (preferably a small one). This approach was applied to a periodically driven thermostat \citep{zhao18}, and shown to successfully differentiate between regular, chaotic and stochastic realizations of time series. However, very long time series, with $n=200,000$, were utilized to demonstrate such phenomenon.

\section{Results}
\label{sect3}

The results are discussed in the following Sections. Table~\ref{tbl1} describes the contents of the accompanying online-only file containing all results.

\begin{table*}
\caption{Contents of the table with time series and PSD properties of the GRBs.}
\label{tbl1}
\centering
\begin{tabular}{clll}
\hline\hline
Column & Column name & Symbol & Description \\
\hline
1  & \texttt{Number} &  Number & Consecutive number of the GRB in the sample (reverse chronological order) \\
2  & \texttt{GRB} & GRB name & Identifying name of the GRB, according to the Swift catalog \\
3  & \texttt{T90} & $T_{90}$ & Duration of the GRB, in seconds \\
4  & \texttt{betaPL} & $\beta_{PL}$ & Exponent $\beta$ of the pure PL fit \\
5  & \texttt{e\_betaPL} & $\Delta\beta_{PL}$ & Uncertainty of the exponent $\beta$ of the pure PL fit \\
6  & \texttt{betaPLC} & $\beta_{PLC}$ & Exponent $\beta$ of the PL plus Poisson noise (PLC) fit \\
7  & \texttt{e\_betaPLC} & $\Delta\beta_{PLC}$ & Uncertainty of the exponent $\beta$ of the PL plus Poisson noise (PLC) fit \\
8  & \texttt{beta1SBPL} & $\beta_{1}$ & Low-frequency exponent $\beta_1$ of the SBPL fit \\
9  & \texttt{e\_beta1SBPL} & $\Delta\beta_{1}$ & Uncertainty of the low-frequency exponent $\beta_1$ of the SBPL fit \\
10 & \texttt{beta2SBPL} & $\beta_{2}$ & High-frequency exponent $\beta_2$ of the SBPL fit \\
11 & \texttt{e\_beta2SBPL} & $\Delta\beta_{2}$ & Uncertainty of the high-frequency exponent $\beta_2$ of the SBPL fit \\
12 & \texttt{Tbreak} & $T_{\rm break}$ & Break time scale of the SBPL fit, in seconds \\
13 & \texttt{e\_Tbreak} & $\Delta T_{\rm break}$ & Uncertainty of the break time scale of the SBPL fit, in seconds \\
14 & \texttt{MVTS} & $\tau$ & Minimum variability time scale, in seconds \\
15 & \texttt{e\_MVTS} & $\Delta\tau$ & Uncertainty of the minimum variability time scale, in seconds \\
16 & \texttt{H} & $H$ & Hurst exponent \\
17 & \texttt{e\_H} & $\Delta H$ & Uncertainty of the Hurst exponent \\
18 & \texttt{HPL} & $H_{PL}$ & Hurst exponent inferred from the index $\beta_{PL}$ \\
19 & \texttt{e\_HPL} & $\Delta H_{PL}$ & Uncertainty of the Hurst exponent inferred from the index $\beta_{PL}$ \\
20 & \texttt{z} & $z$ & Redshift \\
21 & \texttt{Epeak} & $E_{\rm peak}$ & Peak energy of the spectral model, in keV \\
22 & \texttt{e\_Epeak} & $\Delta E_{\rm peak}$ & Uncertainty of the peak energy of the spectral model, in keV \\
23 & \texttt{logLiso} & $\log L_{\rm iso}$ & Logarithm of the peak isotropic luminosity ( $L_{\rm iso}$ in ${\rm erg}\,{\rm s}^{-1}$)\\
24 & \texttt{e\_logLiso} & $\Delta\log L_{\rm iso}$ & Uncertainty of the logarithm of the peak isotropic luminosity\\
\hline
\end{tabular}
\tablecomments{This table is available in its entirety in machine-readable form.}
\end{table*}

\subsection{PSDs}
\label{sect3.1}

\begin{figure*}
\centering
\includegraphics[width=0.49\textwidth]{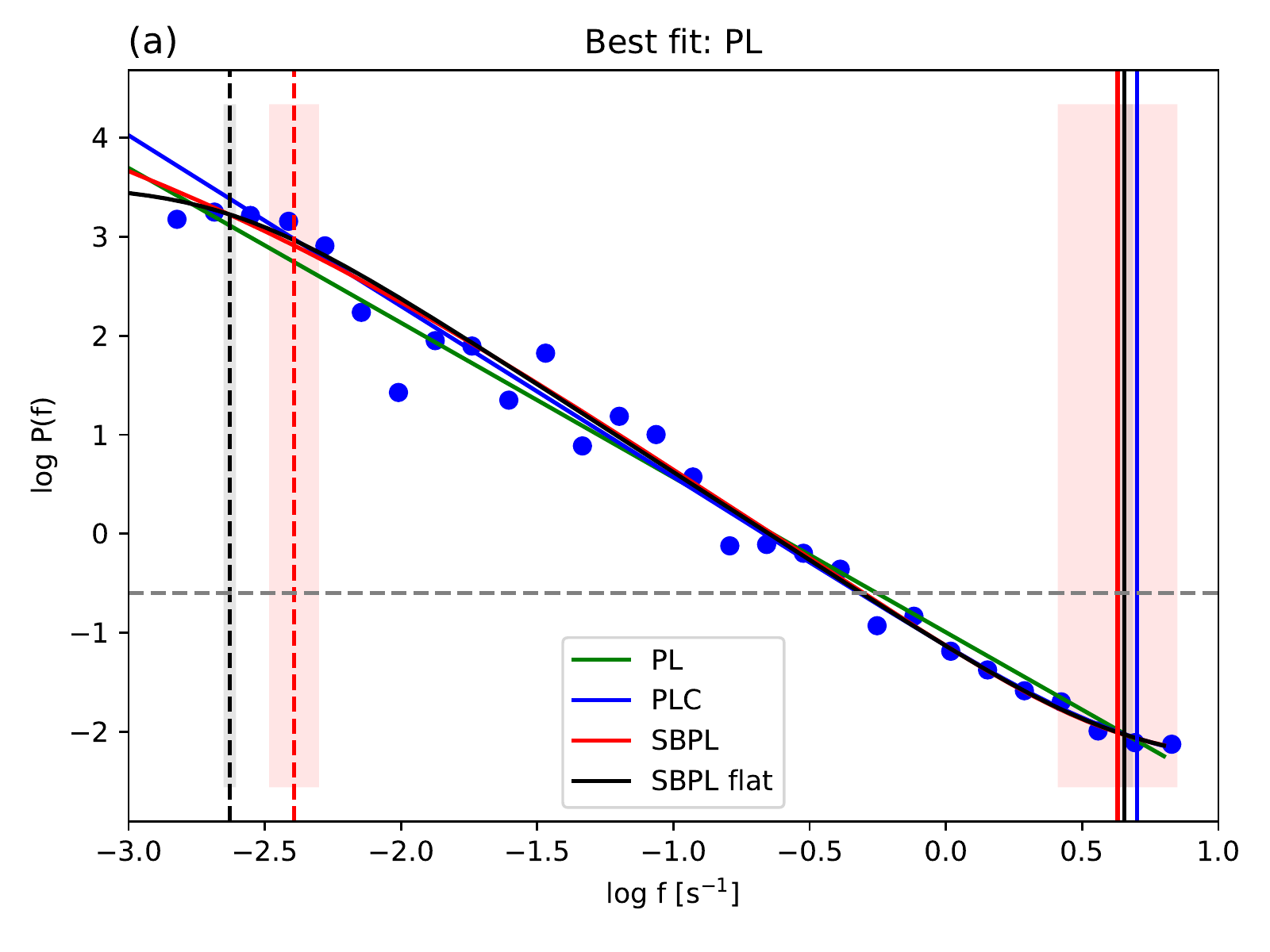}
\includegraphics[width=0.49\textwidth]{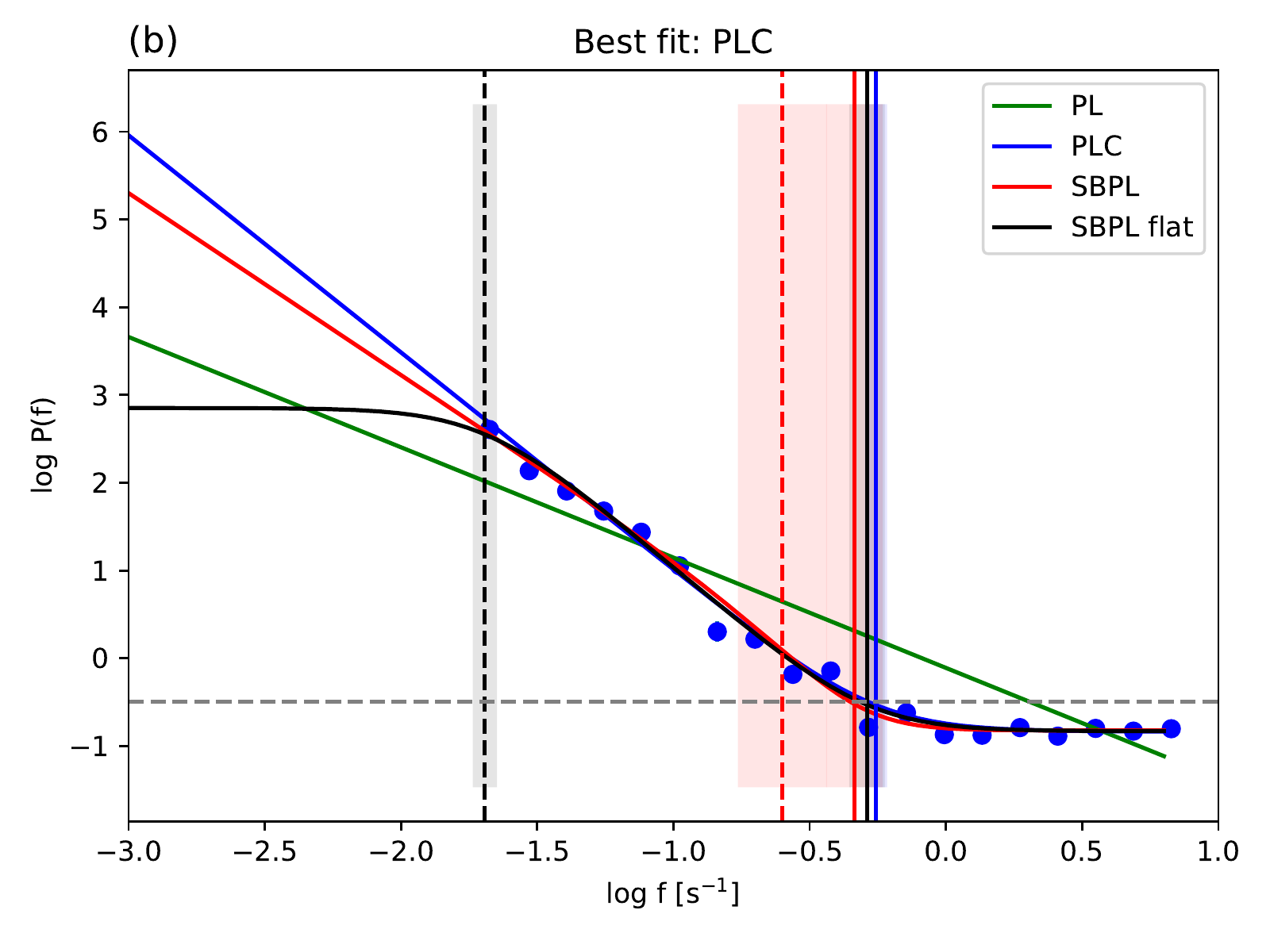}\\
\includegraphics[width=0.49\textwidth]{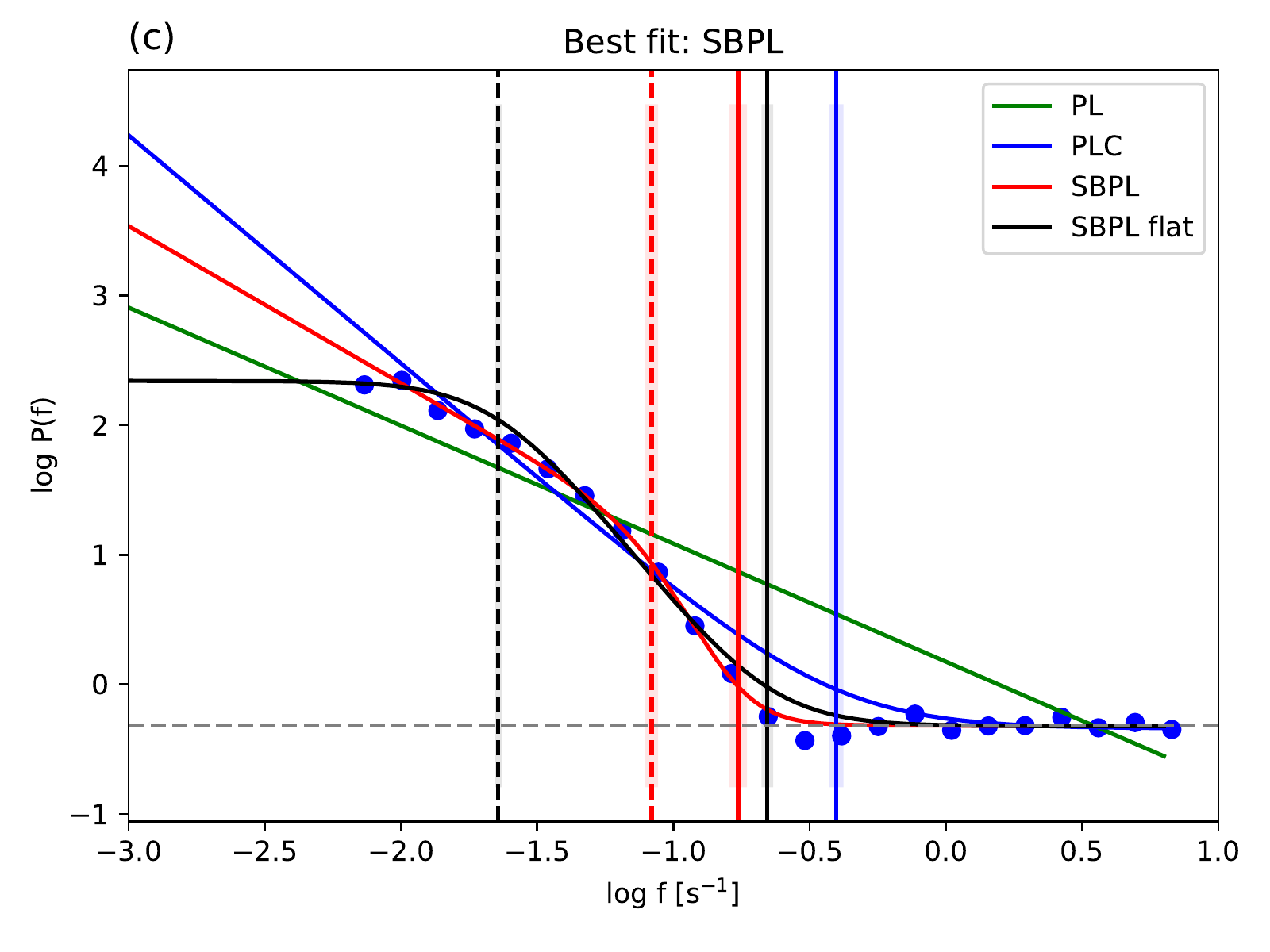}
\includegraphics[width=0.49\textwidth]{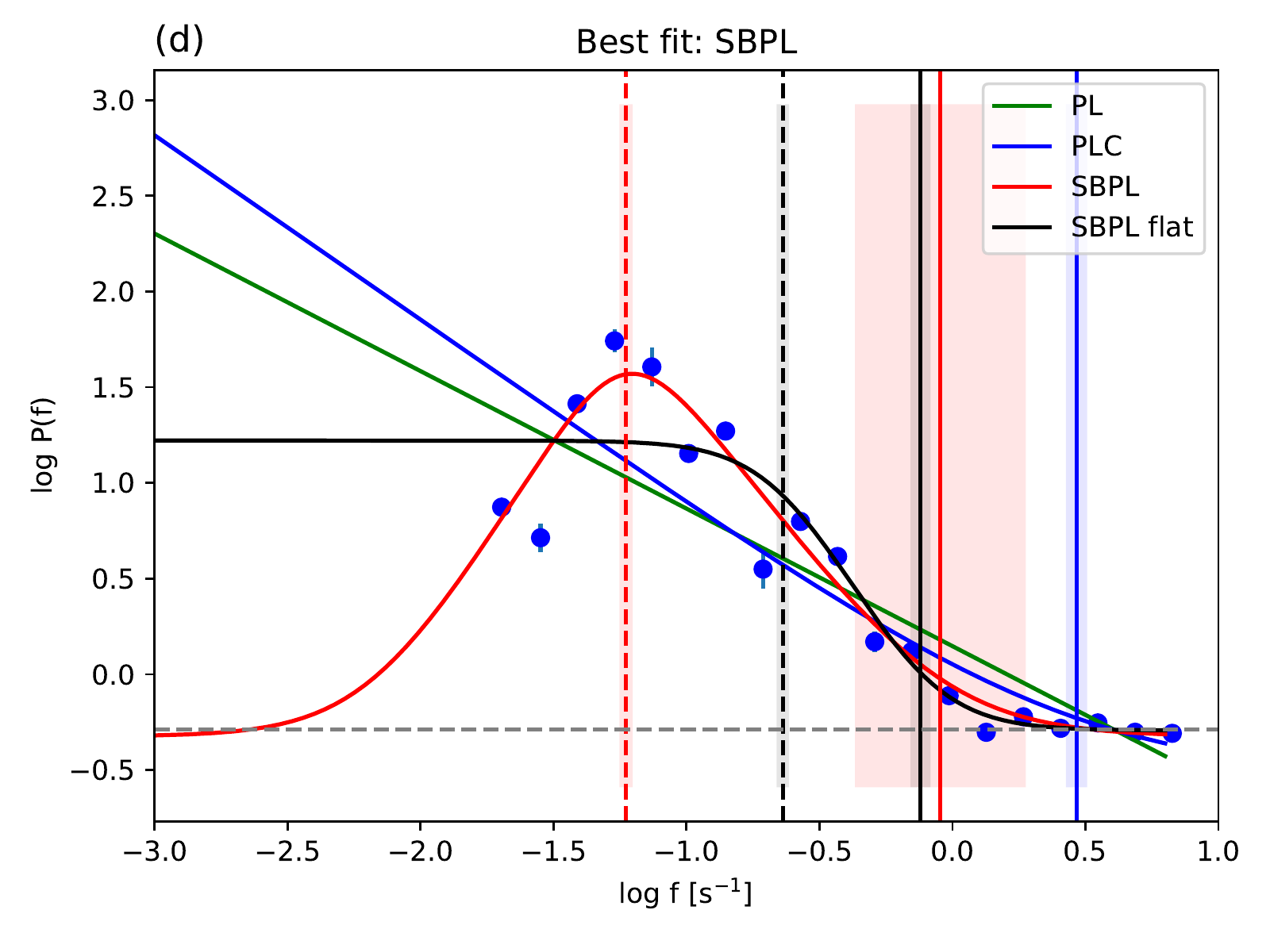}
\caption{Representative PSD forms. (a) GRB 130907A exhibits a pure PL; (b) GRB 120116A---PLC; (c) GRB 070318---SBPL; (d) GRB 190821A---a PSD dominated by a QPO shape. The horizontal gray dashed lines mark the Poisson noise levels inferred from the measurements' errors. The vertical solid lines denote $f_0$: blue---PLC; red---SBPL; black---SBPL with fixed $\beta_1=0$. Vertical dashed lines mark $f_{\rm break}$ of: red---SBPL; black---SBPL with fixed $\beta_1=0$. Widths of the shaded regions symbolize the standard errors of $f_0$ and $f_{\rm break}$.  }
\label{fig0}
\end{figure*}

\begin{figure*}
\centering
\includegraphics[width=\textwidth]{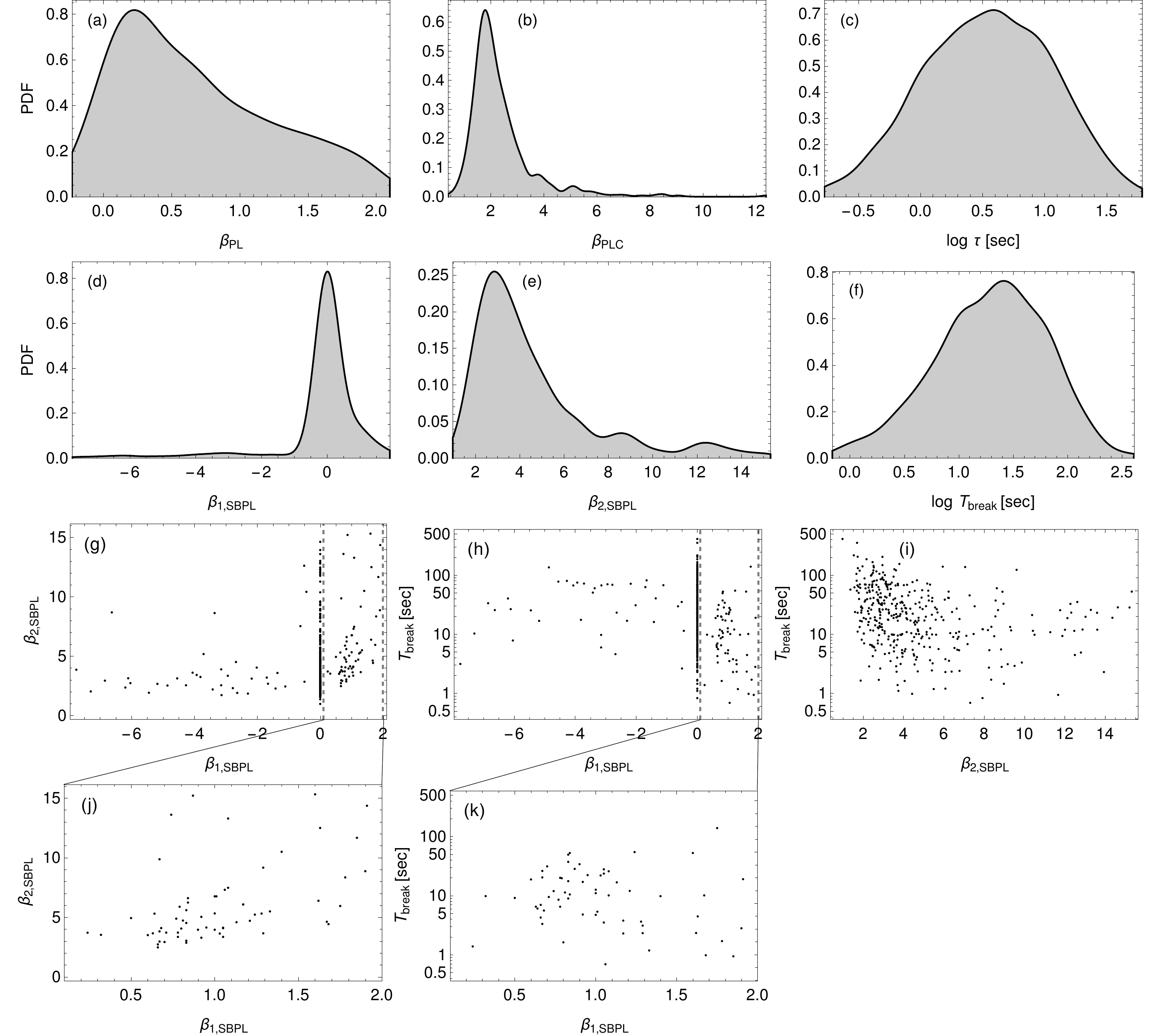}
\caption{Histograms of the $\beta$ indices of the (a) PL and (b) PLC, (c) the MVTS ($>64\,{\rm ms}$) of all applicable cases, the (d) low- and (e) high-frequency indices of the SBPL model, and (f) the break time scales of the SBPL model. The relations between SBPL parameters: (g) $\beta_1-\beta_2$, and $T_{\rm break}$ vs. (h) $\beta_1$ and (i) $\beta_2$. Vertical dashed lines in panels (g) and (h) highlight the regions within which correlation coefficients were calculated; (j) and (k) are the magnifications of the indicated regions. The correlation coefficients and 95\% confidence intervals (CIs) within the regions are (i) $r=-0.26$ (95\% CI: $(-0.35,-0.16)$), (j) $r=0.48$ (95\% CI: $(0.26,0.65)$), and (k) $r=0.12$ (95\% CI: $(-0.13,0.36)$).}
\label{fig1}
\end{figure*}

We were able to obtain meaningful fits of Eqs.~(\ref{eq4})--(\ref{eq6a}) to the PSDs of 1132 GRBs. The best fit was chosen based on the $AIC_c$. As a result, 207 PSDs were modeled best by a pure PL, 548 by a PLC, and 377 yielded SBPL (among which 277 had fixed $\beta_1=0$). 831 MVTS with $\Delta\tau<\tau$ were recorded among the PLC and SBPL cases. Exemplary fits are shown in Fig.~\ref{fig0}. The distributions of the $\beta$, $\beta_1$, $\beta_2$ indices, time scales $\tau$ and $T_{\rm break}$, as well as the scatter plots illustrating the relations between the SBPL parameters, are displayed in Fig.~\ref{fig1}. When the PL model is considered, many GRBs exhibit flat PSDs, $|\beta_{PL}|\lesssim 0.5$, i.e. closely resembling white noise (Fig.~\ref{fig1}(a)), owing to the weakness of the burst and significant Poisson noise contamination. The values extend to $\beta_{PL}\sim 2$, while $\beta_{PLC}$ concentrates around $\beta_{PLC}\sim 2$, with the bulk of it spanning the range $\sim$1--6 (Fig.~\ref{fig1}(b)). In the SBPL model, the indices mostly concentrate around $\beta_{1}\sim 0$ and $\beta_{2}\sim 3$, although there are heavy tails in both distributions, extending to $\beta_{1}<-2$ and $\beta_{2}>8$ (Fig.~\ref{fig1}(d) and (e)). Very steep PSDs (i.e., with high value of $\beta_{2}$) essentially imply no variability on the associated timescales, because the power drops drastically from the conventional PL at lower frequencies to the Poisson noise level at higher frequencies. This means that in these instances there is a sharp cut-off at $T_{\rm break}$ below which variability on shorter timescales is wiped out (excluding the region of Poisson noise domination).

A prominent turnover (i.e., $\beta_{1}$ being very negative) sometimes leads to QPO-like features as in Fig.~\ref{fig0}(d). Often it is a break incorporating just a handful of points in the binned PSD, though. The break time scale $T_{\rm break}$ falls mostly in the range 1--100~s, about an order of magnitude greater than the MVTS (Fig.~\ref{fig1}(c) and (f)). There are moderate or weak correlations between the parameters of the SBPL model (i.e., $\beta_{1}$, $\beta_{2}$, and $T_{\rm break}$; Fig.~\ref{fig1}(g)--(k)). 

It should be emphasized that the Poisson noise levels $C$ obtained by fitting Eqs.~(\ref{eq5})--(\ref{eq6a}) are extremely well correlated ($r=0.98$; 95\% CI: $(0.977,0.982)$) with the estimates inferred from the LC uncertainties via Eq.~(\ref{eq6b}). This means that the statistical fluctuations are the predominant origin of MVTS in the Swift sample, and hence it does not carry physical interpretation regarding the GRB progenitors. In other words, MVTS is the time scale above which the actual signal present in the GRB breaks above the (background or instrumental) noise level.

\subsection{QPOs}
\label{sect3.2}

We searched for QPOs with leading periods $T<T_{100}/3$, i.e. with at least three cycles within the LC. A significance $>3\sigma$ was required, and a visual inspection of the LCs was also done to check for spurious modulations and to ascertain the persistence of the detected phenomenon. We identified 24 QPO candidates with at least one well defined leading period, summarized in Table~\ref{tbl2}. Additionally, in 10 cases we observed a chirping signal, i.e. the leading period evolving in time. In cases when the frequency increases (period decreases) it is called an up-chirp. Similarly, when the frequency decreases (period increases) one encounters a down-chirp.

\begin{deluxetable*}{clcl}
\tabletypesize{\footnotesize}
\tablecolumns{4}
\tablewidth{0pt}
\tablecaption{ Identified QPOs. \label{tbl2}}
\tablehead{
\colhead{Number} & \colhead{GRB name} & \colhead{Period (sec)} & \colhead{Comment}
}
\startdata
6    & GRB200107B & $7.49\pm 1.16$; $11.40\pm 1.57$ & harmonics, $2:3$ \\
34   & GRB190821A & $8.20\rightarrow 5.28$ & up-chirp \\
75   & GRB190103B & $5.17\pm 0.76$ & constant \\
102  & GRB180823A & $19.12\pm 3.19$ & constant \\
122  & GRB180626A & $4.58\pm 0.33$; $5.70\pm 0.54$ & harmonics, $4:5$ \\
190  & GRB170823A & $2.96\rightarrow 11.58$ & down-chirp \\
212  & GRB170524B & $2.1\rightarrow 2.8$ & down-chirp \\
232  & GRB170205A & $6.86\pm 0.80$ & constant \\
250  & GRB161202A & $24.27\rightarrow 16.25$ & up-chirp \\
251  & GRB161129A & $3.83\rightarrow 6.95$ & up-chirp \\
252  & GRB161117B & $3.82\pm 0.52$ & constant \\
272  & GRB160824A & $3.05\pm 0.56$; $5.37\pm 0.81$; $9.43\pm 1.51$ & harmonics, $4:7:12^*$ \\
455  & GRB140730A & $12.32\pm 1.95$ & constant \\
462  & GRB140709B & $20.90\pm 2.00$; $41.54\pm 4.30$ & harmonics, $1:2$ \\
470  & GRB140619A & $8.87\pm 0.99$; $13.10\pm 1.85$; $32.34\pm 3.86$ & harmonics, $6:15:22^*$ \\
496  & GRB140323A & $5.49\pm 0.98$; $21.31\pm 2.91$ & harmonics, $1:4$ \\
551  & GRB130812A & $2.26\pm 0.40$ & constant \\
618  & GRB121209A & $9.89\rightarrow 7.57$ & up-chirp \\
622  & GRB121125A & $4.29\pm 0.73$; $8.48\pm 1.00$ & harmonics, $1:2$ \\
632  & GRB121014A & $16.70\pm 1.87$ & constant \\
701  & GRB120116A & $8.16\pm 0.96$ & constant \\
756  & GRB110422A & $5.46\rightarrow 3.89$ & up-chirp \\
777  & GRB110207A & $6.26\pm 0.74$ & constant \\
783  & GRB110107A & $5.48\rightarrow 3.46$ & up-chirp \\
805  & GRB100924A & $20.18\rightarrow 5.14$ & up-chirp \\
914  & GRB090709A & $8.02\pm 0.67$; $9.80\pm 0.91$ & harmonics, $4:5$ \\
945  & GRB090404 & $10.94\pm 0.86$ & constant \\
963  & GRB090102 & $7.64\pm 1.07$ & constant \\
1007 & GRB080810 & $6.70\pm 0.60$; $9.15\pm 0.85$; $12.67\pm 0.81$ & harmonics, $2:3:4$ \\
1098 & GRB070911 & $4.97\pm 0.75$; $16.50\pm 2.08$ & harmonics, $3:10$ \\
1127 & GRB070508 & $2.14\pm 0.26$; $4.43\pm 0.87$ & harmonics, $1:2$ \\
1185 & GRB060906 & $4.77\pm 0.68$ & constant \\
1324 & GRB050418 & $14.70\rightarrow 4.76$ & up-chirp \\
1335 & GRB050306 & $27.97\pm 3.93$ & constant \\
\enddata
\tablecomments{Approximately constant leading periods are given with corresponding uncertainties (indicated with the '$\pm$' sign). Period ranges of the chirping signals are indicated with arrows, '$\rightarrow$', showing the direction of period evolution. For the harmonics, the closest integer ratios are provided. \\$^*$These high-order ratios might as well be spurious, or be obscured due to uncertainties.}
\end{deluxetable*}

There are 13 GRBs with one, prominent QPO with a constant leading period; 8 GRBs with two coexisting QPOs; and 3 GRBs with three coexisting QPOs. In the latter two cases we refer to them as 'harmonics'. The periods range from 2.14 to 41.54~s. In Table~\ref{tbl2}, closest integer ratios (resonances) of the detected periods are also proposed. Most of the double-QPO cases are of low or moderate orders (except for GRB 070911, which has a $3:10$ ratio), with the $1:2$ ratio occurring in three instances. Triple-QPOs seem to exhibit high-order, likely spurious ratios, except for GRB 080810, which yields the second lowest possible ratio of $2:3:4$. Among the chirping signals, 8 are up-chirps and only 2 are down-chirps.

\begin{figure*}
\centering
\includegraphics[width=0.8\textwidth]{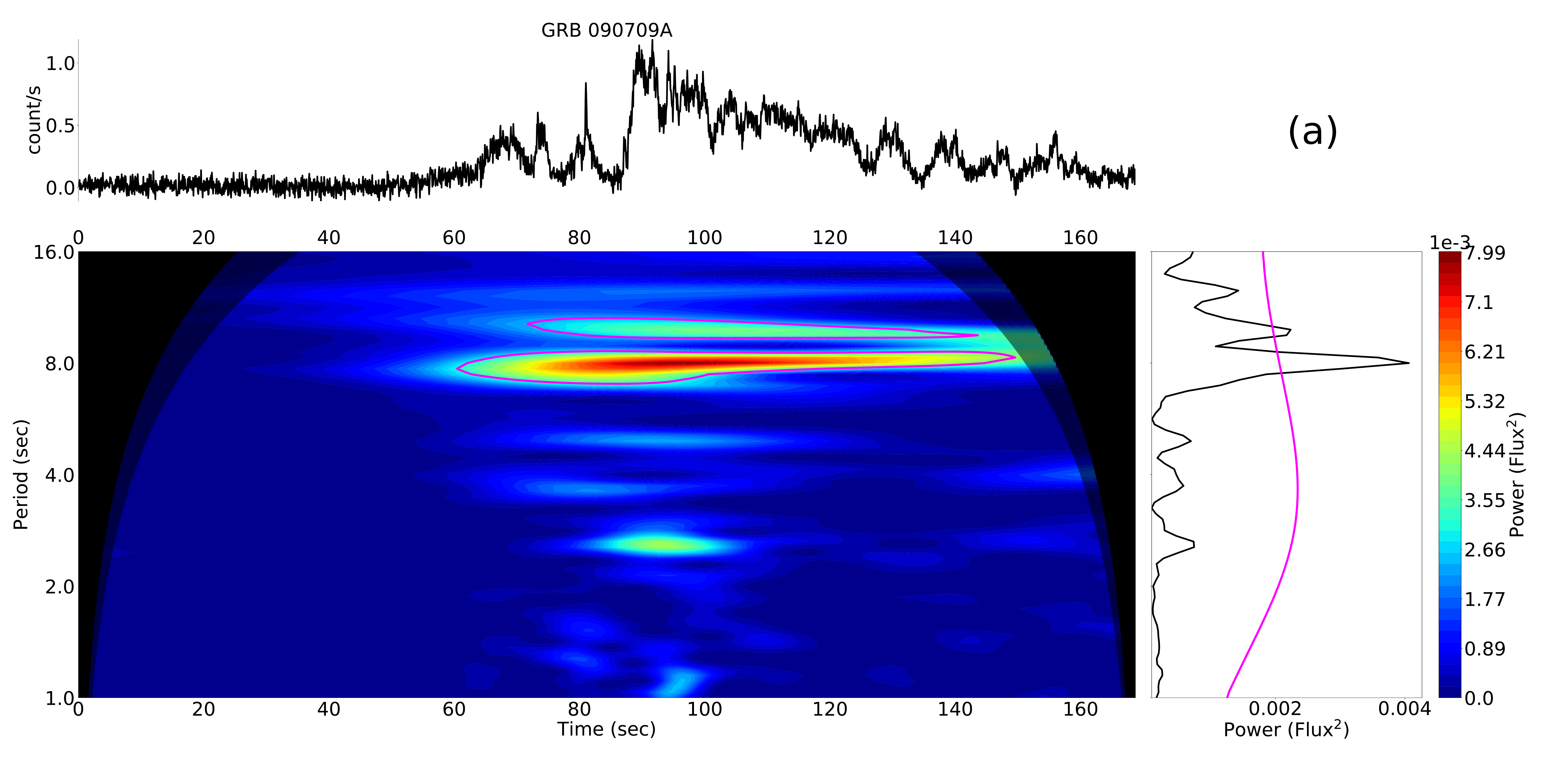}\\
\includegraphics[width=0.8\textwidth]{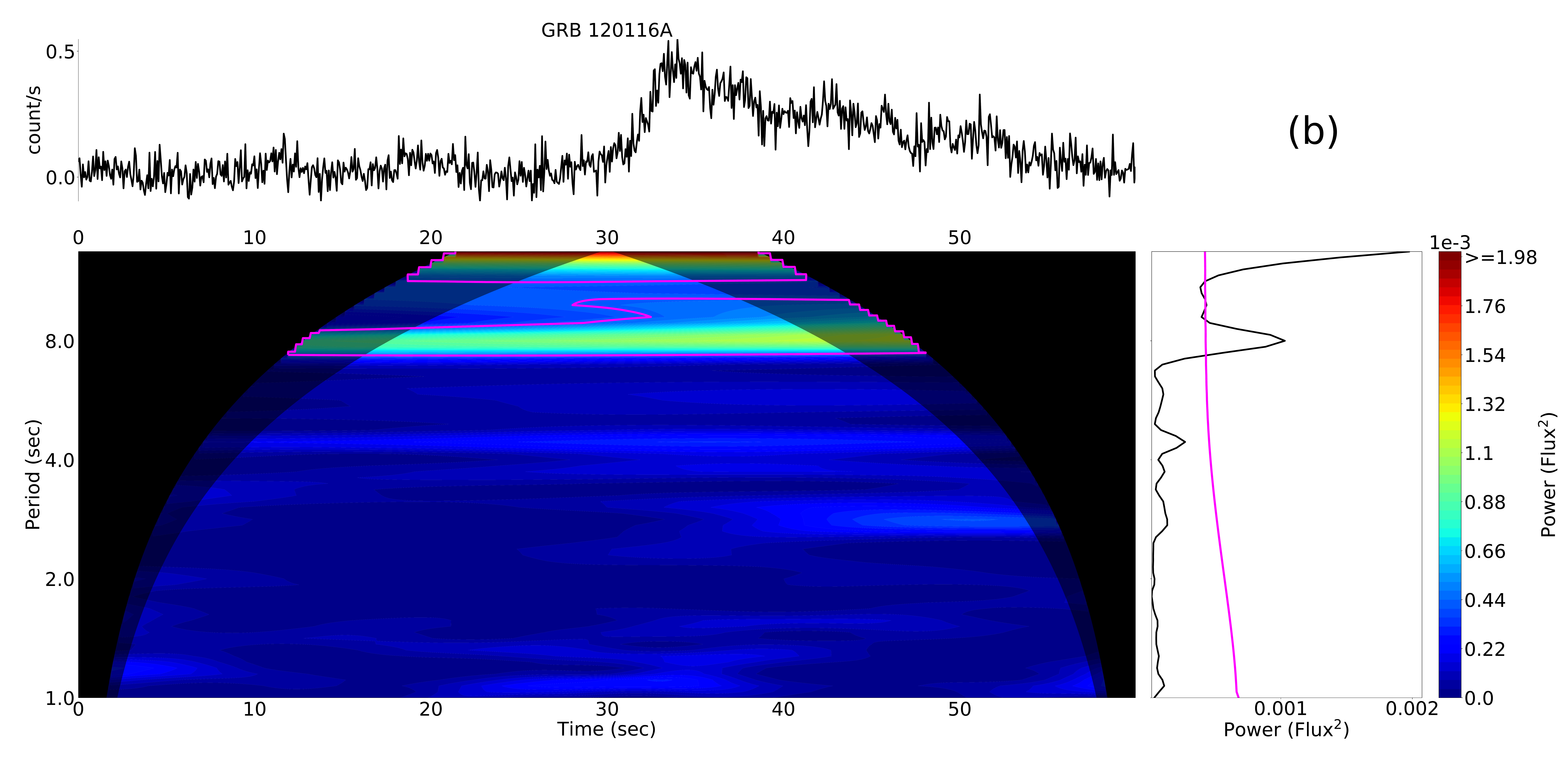}
\caption{ (a) Wavelet scalogram of GRB 090709A. There is a significant (>3$\sigma$) QPO, with a leading period $\sim 8$~sec, persistent through most of the LC. Another, slightly shorter component is visible at a period $\sim 9.8$~sec. (b) GRB 120116A exhibits very similar features. }
\label{scalogram914}
\end{figure*}

Fig.~\ref{scalogram914}(a) shows the scalogram for GRB 090709A, a source with an 8~s quasiperiodicity, overlaid on a FRED-like pulse, already ambiguously reported \citep{markwardt09,deluca10,cenko10}. Despite being speculated to be an SGR, it is almost surely a GRB. We confirm this QPO on a 3$\sigma$ significance level, and obtain a leading period of 8.02~s, persistently spanning almost 100~s of the LC. Moreover, the scalogram reveals another QPO, with a slightly higher period of $\sim$9.8~s, lasting for about 70~s, and contemporary with the 8~s QPO. The ratio of the periods is close to a $4:5$ (or $5:6$) resonance---a moderate order. \citet{ziaeepour11} showed that invoking a precession of a strong external magnetic field (present in case of stars believed to be progenitors of long GRBs) might lead to an oscillating behavior in the prompt LC. The overall nature of the two QPOs present in this GRB is unclear, though. We obtained a very similar picture for GRB 120116A, which has the same overall FRED-like shape with an 8\,s QPO overlaid (Fig.~\ref{scalogram914}(b)). The QPOs in both GRBs are thence likely to be a result of the same mechanism and conditions at the emission site.

\begin{figure*}
\centering
\includegraphics[width=0.8\textwidth]{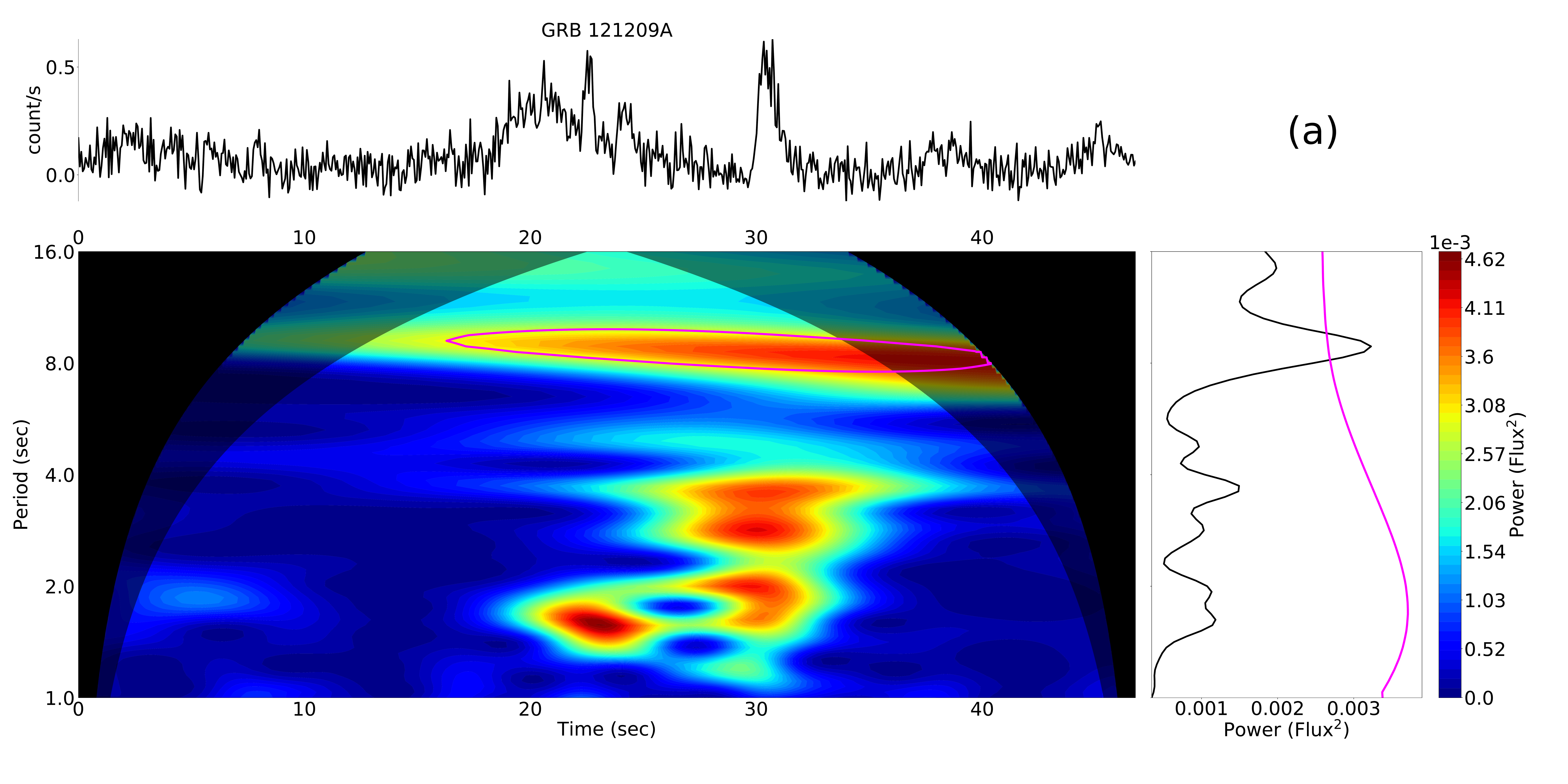}\\
\includegraphics[width=0.8\textwidth]{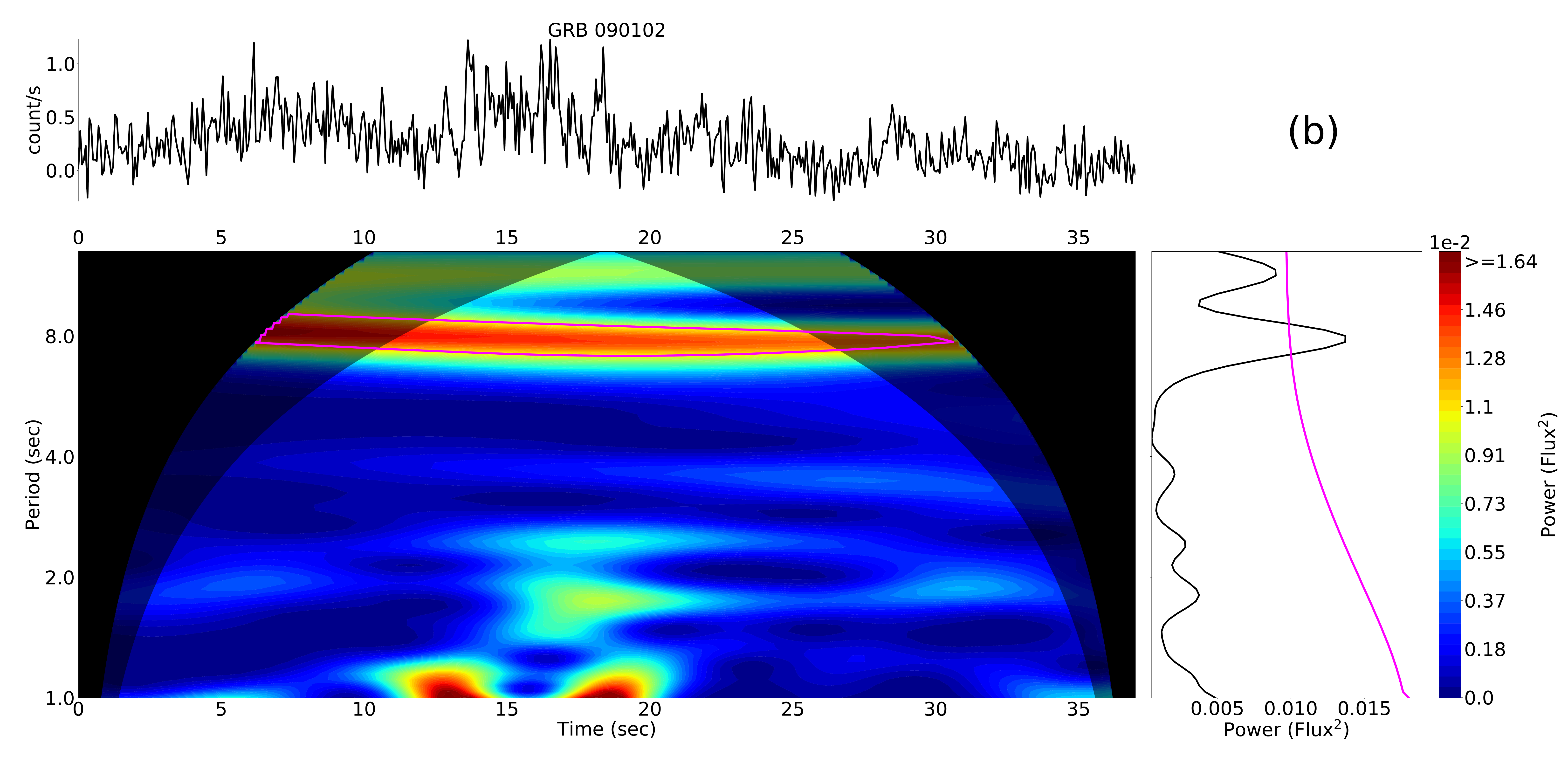}\\
\includegraphics[width=0.8\textwidth]{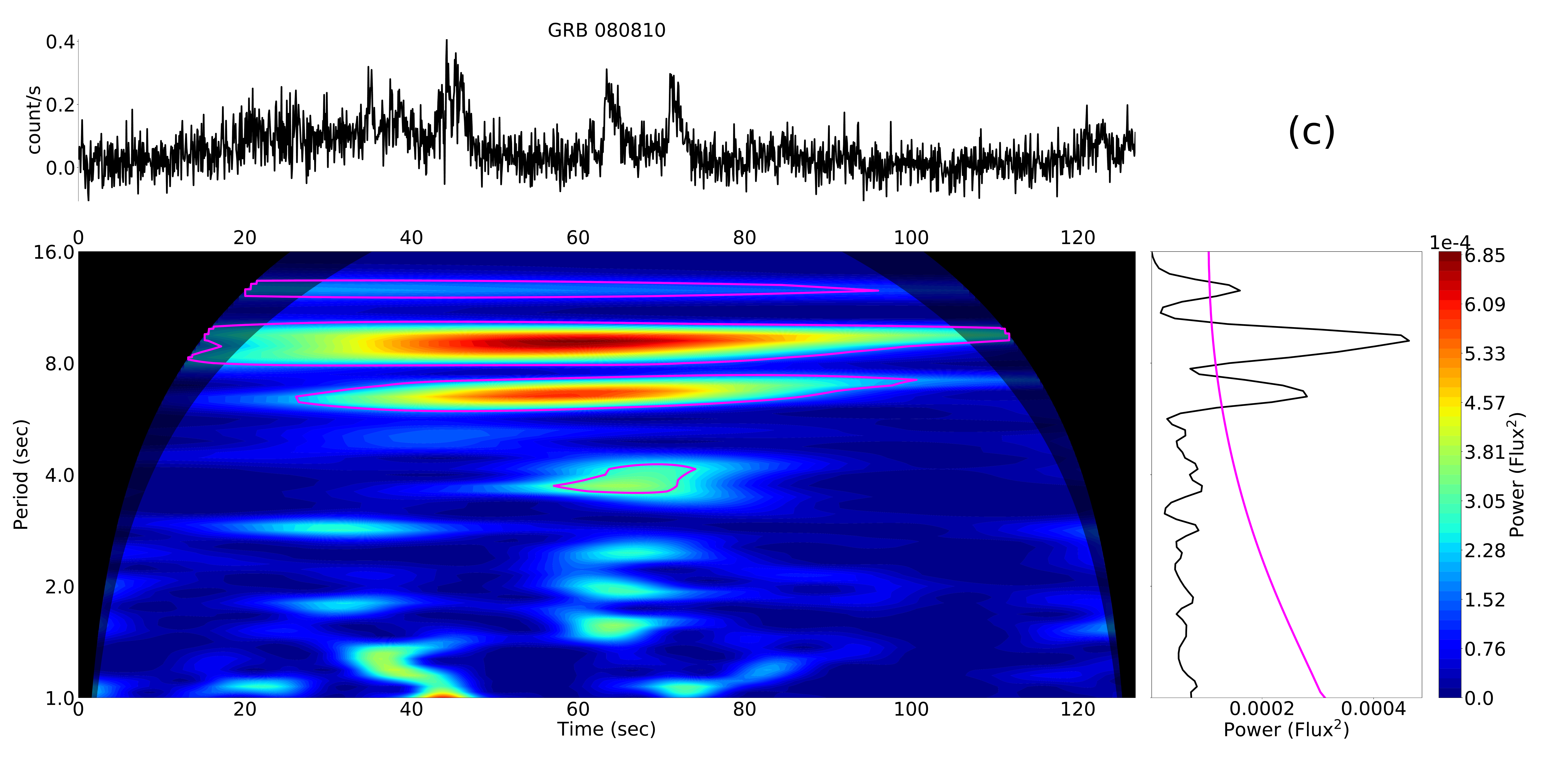}
\caption{ Exemplary wavelet scalograms of (a) an up-chirp in GRB 121209A, (b) a constant leading period in GRB 090102, and (c) a $2:3:4$ resonance in GRB 080810. }
\label{scalograms_examples}
\end{figure*}

Fig.~\ref{scalograms_examples} shows examples of novel detections of a slightly chirping signal (Fig.~\ref{scalograms_examples}(a)), another constant leading period (Fig.~\ref{scalograms_examples}(b)), and a case of harmonics remaining in an apparent $2:3:4$ resonance (Fig.~\ref{scalograms_examples}(c)).

\subsection{Hurst exponents}
\label{sect3.3}

\begin{figure*}
\centering
\includegraphics[width=\textwidth]{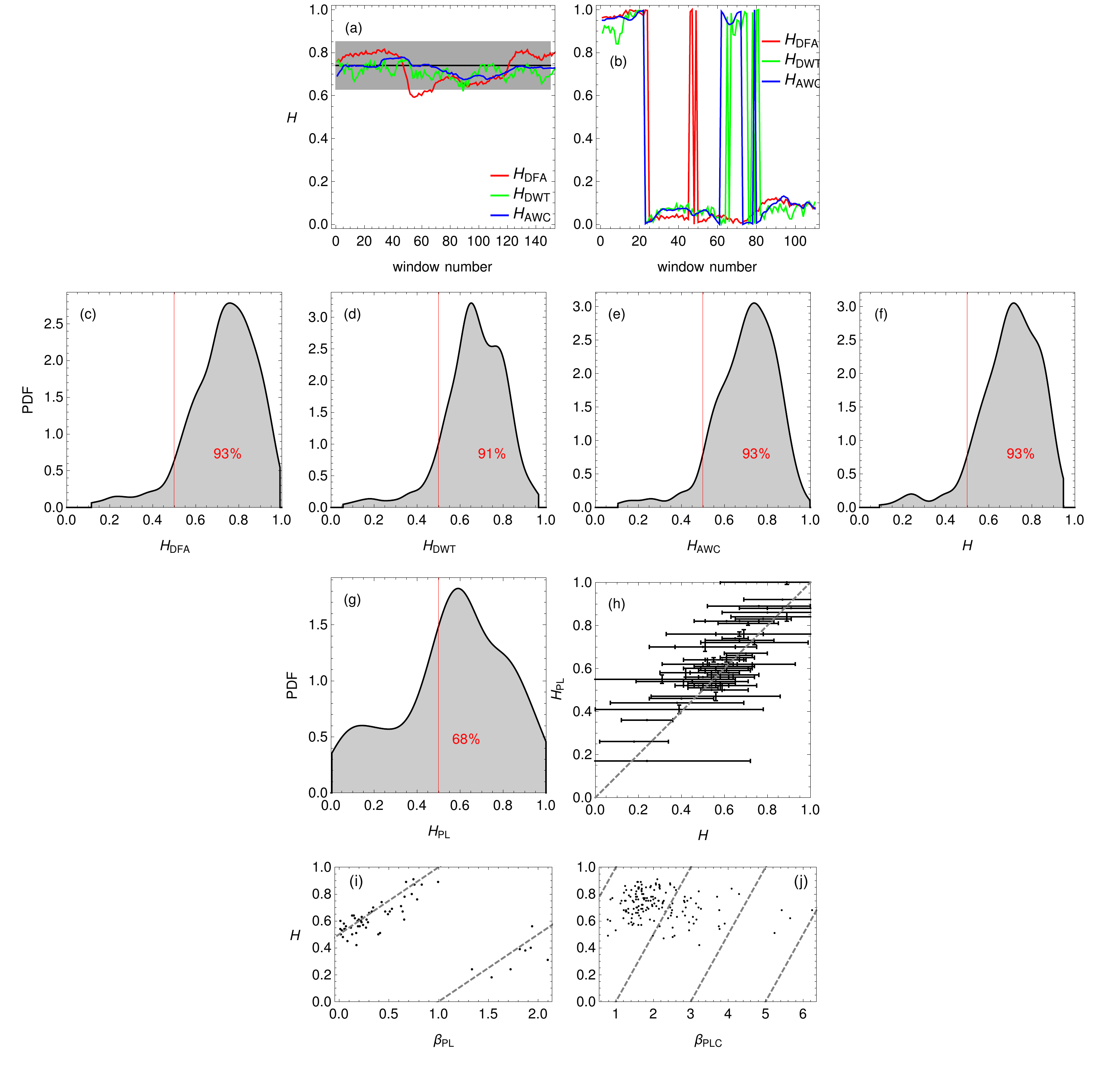}
\caption{ (a)--(b) Time evolution of $H$. (a) The estimates for GRB 191001B are consistent with each other. The global value of $H=0.74$ is indicated with a horizontal black line, and the gray region around it marks the uncertainty of $\Delta H=0.11$. (b) For GRB 180111A no unambiguous estimate can be obtained. (c)--(f) Distributions of the Hurst exponents: (c)--(e) obtained with different methods (DFA, DWT, AWC, respectively), and (f) the final values, $H$, being the mean of the three. Vertical red lines denote $H=0.5$, and the percentage of cases $H>0.5$ is indicated in each panel. (g) Distribution of the Hurst exponents obtained directly from the values of $\beta_{PL}$. (h) Relation between the $H$ obtained with other methods, and $H_{PL}$ obtained from the values of $\beta_{PL}$. Diagonal dashed line marks the identity relation. (i)--(j) Hurst exponents obtained directly from the PL indices in case of a pure PL and PLC, respectively. Inclined dashed lines depict the theoretical relations. In (i), the overall correlation is $r=0.9$ (along the theoretical line when considered continuous; 95\% CI: $(0.84,0.94)$), while in (j) $r=-0.18$ (95\% CI: $(-0.33,-0.02)$), and does not follow well the predicted values. }
\label{fig4}
\end{figure*}

To estimate $H$, the algorithms from Sect.~\ref{sect2.4} (DFA, DWT, AWC) were utilized. To obtain robust estimates, first were selected only those GRBs for which all three $H$ estimates were consistent with each other, within the standard errors. Next, a time evolution of the three $H$ values were investigated, and only cases with no sudden jumps between $H\sim 0$ and $H\sim 1$ were kept. Given a time series with length $n$, it was divided into sliding windows of size $\floor*{n/2}$, resulting in $\ceil*{n/2}$ such chunks. To each, the three algorithms were applied, and hence provided the time evolution of $H$. Examples of this procedure are shown in Fig.~\ref{fig4}(a) and (b). This eventually led to 335 estimates of $H$, whose distributions are displayed in Fig.~\ref{fig4}(f). Over 90\% of GRBs are characterized by $H>0.5$, meaning they possess long-term memory. This is highly consistent with the overall shape of most LCs, which are comprised of one or more FRED-like pulses. Overall, a pulse by itself is persistent: the initial rise lasts for a prolonged period of time (longer than the sampling time step), and is followed by a prolonged decay, i.e. a trend is present in an LC, leading to $H>0.5$. Therefore, when on the rising side of the pulse, one can expect that the rise will continue, and when on the decaying part, one shall expect it to further continue its decay. A low signal-to-noise ratio can, however, allow the statistical fluctuations in form of white or otherwise anticorrelated noise to dominate, hence leading to $H<0.5$ (short-term memory). On the other hand, some of the few cases yielding $H<0.5$ also exhibit pronounced pulses. $H$ can be applied to both stationary and nonstationary processes, and both types can exhibit short- and long-term memory. In many GRBs the variability is clearly nonstationary, but the governing process may as well be antipersistent (i.e., be characterized by $H<0.5$). It therefore follows that the distinction $H\lessgtr 0.5$ is not trivially connected merely with the shape of the pulses or the signal-to-noise ratio of the LCs.

Additionally, since there is a theoretical linear relation between $H$ and the PL index $\beta_{PL}$, for the 187 GRBs with a PL PSD the $H$ were extracted directly from the $\beta_{PL}$ values. Their distribution is displayed in Fig.~\ref{fig4}(g), while Fig.~\ref{fig4}(h) demonstrates that these values are consistent with the $H$ obtained with the other three algorithms. Indeed, these latter $H$ exhibit a very high correlation with $\beta_{PL}$, in perfect agreement with the theoretical predictions ($r=0.94$, 95\% CI: $(0.92,0.96)$; Fig.~\ref{fig4}(i)). This is, however, not the case when similar inference is performed using the indices $\beta_{PLC}$ from the PLC model (Fig.~\ref{fig4}(j)): there is a weak anticorrelation between the two, and the obtained $H$ values do not follow the theoretical predictions at all. This seems to be the fault of the Poisson noise contaminating the signal, as the $H$ extraction algorithms (DFA, DWT, AWC) treat the time series as a whole, so the random fluctuations obscure the self-affinity that the algorithms rely on. A meaningful inference of $H$ from any signal is hence a subtle matter. 

Finally, we note there is nearly no correlation between $H$ and the parameters of the SBPL model ($\beta_1,\,\beta_2,\,T_{\rm break}$). Recall that no conditions were imposed on the signal-to-noise ratio of the GRBs; we aimed to analyze as much of the Swift catalog as was technically possible.

\subsection{The $\mathcal{A}-\mathcal{T}$ plane}
\label{sect3.4}

\begin{figure*}
\centering
\includegraphics[width=\textwidth]{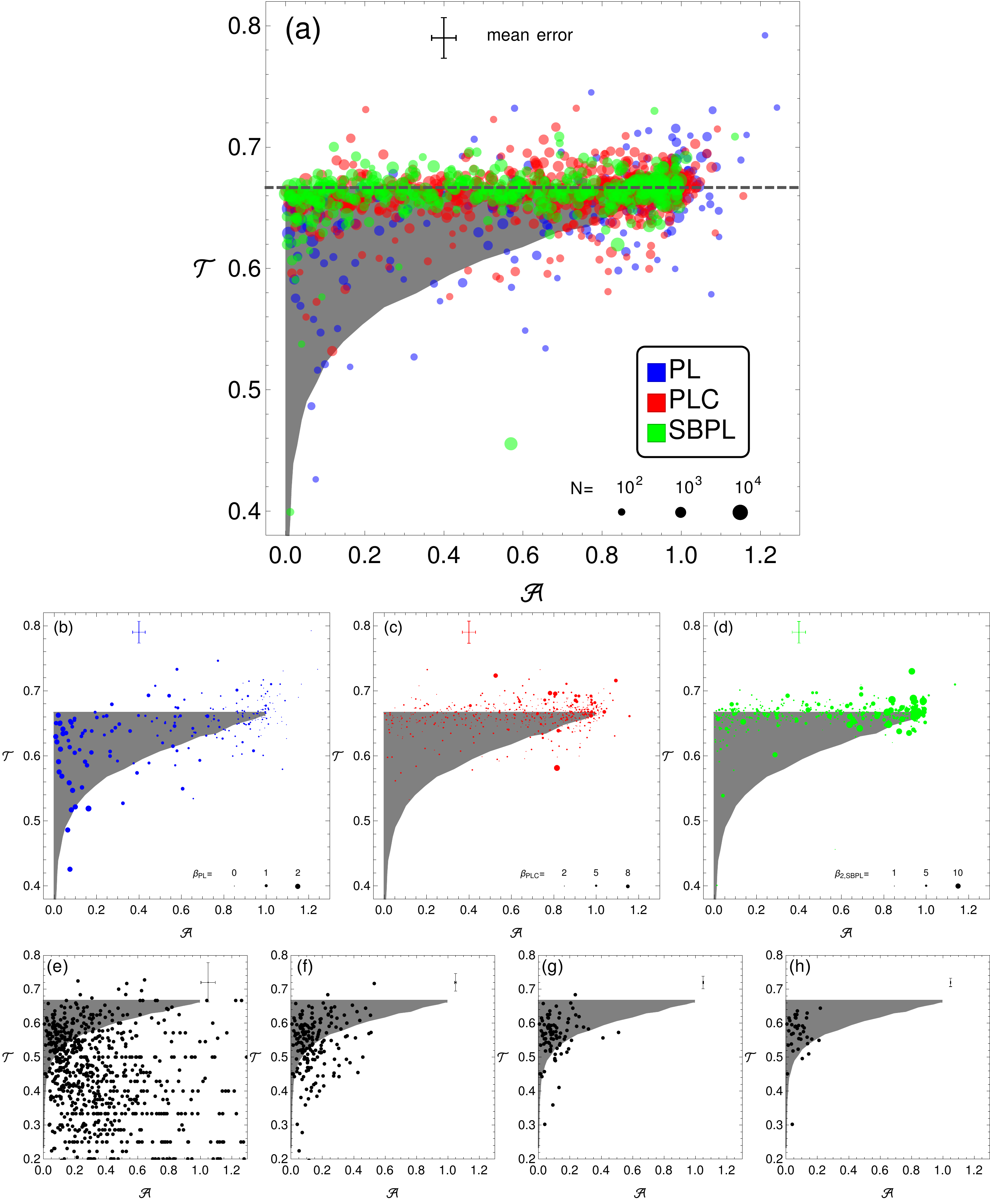}
\caption{ The $\mathcal{A}-\mathcal{T}$ plane. The mean errors are represented in the upper parts of the plots. (a) The point size is proportional to the logarithms of the length of the time series, and indicated in the lower right corner. The gray area is the region of availability for PLC type of PSDs (cf. Fig.~\ref{fig_AT}). (b)--(d) The sizes of the points indicate the values of index $\beta$, as indicated in the lower right corner of each panel. (e)--(h) The $(\mathcal{A},\mathcal{T})$ locations of the LCs binned according to the MVTS values. In panel (e) locations of all such binned LCs are represented, while in (f), (g), and (h) displayed are binned LCs with at least 50, 100, and 200 points, respectively. }
\label{fig_AT_results}
\end{figure*}

The $\mathcal{A}-\mathcal{T}$ plane is linked with the Hurst exponents, as well as the PSD form. In Fig.~\ref{fig_AT_results}(a) displayed are the $(\mathcal{A},\mathcal{T})$ locations of the 1150 GRBs in the 64\,ms binning. The gray area in the background is the region of availability of the PLC, i.e. above the pure PL line (red points in Fig.~\ref{fig_AT}) and below the line $\mathcal{T}=\nicefrac{2}{3}$ (highlighted with a gray dashed line in Fig.~\ref{fig_AT_results}(a)). The size of the point is proportional to the logarithm of the number of measurements $n$ in the LC. There is essentially no prominent correlation between $n$ and both $\mathcal{A}$ and $\mathcal{T}$.

Figures~\ref{fig_AT_results}(b)--(d) display the $\mathcal{A}-\mathcal{T}$ plane as well, but with the size of the points indicating the index $\beta$ of the best-fit PSD. In case of pure PL (Fig.~\ref{fig_AT_results}(b)) the relation is consistent with Fig.~\ref{fig_AT}, i.e. steeper PSDs are located at lower values of $\mathcal{A}$ and---to some extent---lower values of $\mathcal{T}$ as well. However, in case of PLC and SBPL models (Fig.~\ref{fig_AT_results}(c) and (d), respectively) the situation seems to be reverted, with the steepest PSDs crowding near the white noise point $(1,\nicefrac{2}{3})$. Note that very steep PSDs (i.e., with $\beta\gtrsim 4-5$) imply virtually no, or very little, variability on the associated time scales (cf. Sect.~\ref{sect3.1}). The Poisson noise dominates such PLC cases, and a combination of white noises at two different power levels (at time scales $\lesssimgtr T_{\rm break}$, since $T_{\rm break}\approx\tau$ in such instances) occurs in SBPL, especially when $\beta_{1}\approx 0$. After excluding these extremely steep instances, there is no correlation between $\beta$ and both $\mathcal{A}$ and $\mathcal{T}$. Finally, there are strong anticorrelations between $\mathcal{A}$ and $\log f_0$ ($r=-0.77$ for PLC; 95\% CI: $(-0.80,-0.73)$, and $r=-0.76$ for SBPL; 95\% CI: $(-0.80,-0.71)$), confirming that the level of Poisson noise contaminating the LCs primarily determines their locations in the $\mathcal{A}-\mathcal{T}$ plane.

To circumvent this, the LCs were binned according to the MVTS (effectively smoothing the LCs), and the resulting $(\mathcal{A},\mathcal{T})$ locations are shown in Fig.~\ref{fig_AT_results}(e). This picture is completely random, since many binned LCs turned out to contain only a handful of points. Therefore, in Figs.~\ref{fig_AT_results}(f)--(h) are shown only those binned LCs with at least 50, 100, and 200 points, respectively. The longer the binned LCs, the more consistent their locations are with the region of availability of PLC models.

It is therefore crucial to highlight the importance of sufficient length of a time series for calculating its location in the $\mathcal{A}-\mathcal{T}$ plane robustly. While the values of $(\mathcal{A},\mathcal{T})$ in principal can be computed and used to characterize any experimental time series (even extremely short ones), when dealing with stochastic processes, very short realizations will give essentially a random outcome. Consider, e.g., a realization of white noise with $n$ values. It has an expected number of turning points $\mu_T = \nicefrac{2}{3}(n-2)$, and standard deviation $\sigma_T=\sqrt{(16n-29)/90}$ \citep{kendall}. The distribution of $T$ (for a fixed $n$) will tend to a Gaussian parametrized by $(\mu_T,\sigma_T)$. For $n\in\{6,25,50,1000\}$, these are (rounded to the nearest integer): $(\mu_T,\sigma_T)\in\{(3,1), (15,2), (32,3), (665,13)\}$. Translating to $\mu_\mathcal{T}=\mu_T/n$, $\sigma_\mathcal{T}=\sigma_T/n$, one gets $\{(0.5, 0.17), (0.6, 0.08), (0.64, 0.06), (0.665, 0.013)\}$. In other words, the expected value of $\mathcal{T}$ and its standard deviation are asymptotically equal to $\nicefrac{2}{3}$ and zero, respectively, but for short time series $\sigma_T$ (or $\sigma_\mathcal{T}$) can constitute a substantial fraction of $\mu_T$ (or $\mu_\mathcal{T}$). For instance, a time series with only $n=6$ will have 33\%, 42\%, and 17\% chance of yielding $T=2,3,4$, respectively.

Such trend is indeed observed in Figs.~\ref{fig_AT_results}(e)--(h): the longer the binned LCs, the more constrained to the region of availability they are. Therefore, the Poisson noise is a serious obstacle in analyzing the variability of astronomical time series, GRBs in particular. Recall that, according to Sect.~\ref{sect3.1}, the MVTS does not bear any physical meaning (at least for the Swift sample investigated herein), as it very strongly depends on the Poisson noise level inferred from the individual measurements' uncertainties. Finally, given all the above considerations, the $\mathcal{A}-\mathcal{T}$ plane, while potentially useful in classifying LCs \citep[cf. ][]{zywucka20,tarnopolski20b}, does not hint at any clustering of long GRBs into more than one group (other than an overconcentration of flat PSDs at $(1,\nicefrac{2}{3})$, likely owing to low signal-to-noise ratio), consistent with the findings of \citet{jespersen20}.

Finally, the coarse graining (Eq.~(\ref{eq60})) was applied to the 64\,ms binned LCs, using $d\in\{1,\ldots,20\}$, with the intent to possibly obtain separated clusters for some particular value of $d$. Such an approach was successful in case of economic and physiological data \citep{zunino17} with small values of $d$. For the GRBs herein, though, we mostly observe variations of Fig.~\ref{fig_AT_results}(a) for small $d$, and higher $d$ (leading to relatively short coarse-grained sequences) resembling Fig.~\ref{fig_AT_results}(e) when the resulting time series are too short. We therefore again do not obtain any clustering of long GRBs in the $\mathcal{A}-\mathcal{T}$ plane.

\subsection{The $E_{\rm peak}^{\rm rest}-\beta$ relation}
\label{sect3.5}

\begin{figure}
\centering
\includegraphics[width=\columnwidth]{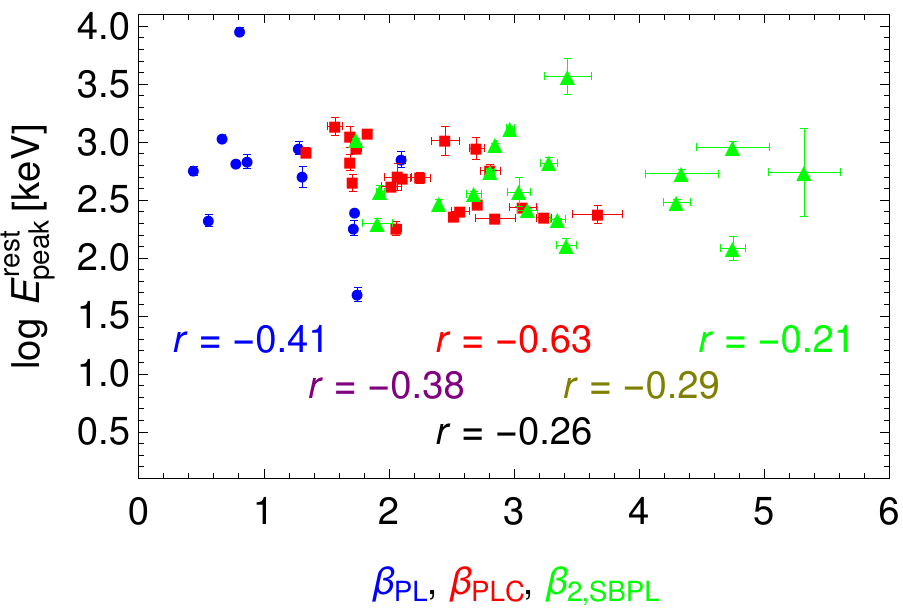}
\caption{ The relation between $E_{\rm peak}^{\rm rest}$ and the $\beta$ indices obtained from PL (blue circles), PLC (red squares), and SBPL models (green triangles), and correlation coefficients highlighted with respective colors, where purple refers to the joint set $\beta = \beta_{PL} \cup \beta_{PLC}$, dirty yellow to $\beta = \beta_{PLC} \cup \beta_{2}$, and black to $\beta = \beta_{PL} \cup \beta_{PLC} \cup \beta_{2}$. }
\label{fig5.3}
\end{figure}

\citet{dichiara16} studied 123 GRBs observed by various instruments. They found a statistically significant anticorrelation ($r=-0.54$, 95\% CI: $(-0.65,-0.41)$, using their published data) between the rest-frame peak energy ($\log E_{\rm peak}^{\rm rest}$) and the PL index (denoted by them with $\alpha$; we continue using the symbol $\beta$ hereinafter). We gathered the $E_{\rm peak}$ values of the \textsc{Band}, \textsc{Comp}, and \textsc{Sbpl}\footnote{Note this \textsc{Sbpl} is in a different context than the PSD from Eq.~(\ref{eq6}).} spectral fits \citep{kaneko06,gruber} from the Fermi/GBM catalog\footnote{\url{https://heasarc.gsfc.nasa.gov/W3Browse/fermi/fermigbrst.html}} \citep{gruber,kienlin,kienlin20,bhat} by cross-matching the spatio-temporal localizations of the Swift and Fermi GRBs, and complemented them with redshift measurements when available. The specific $E_{\rm peak}$ values for the common GRBs were chosen based on the \texttt{flnc\_best\_fitting\_model} entry from the Fermi catalog. We consider $\beta_{PL}$, $\beta_{PLC}$, and $\beta_{2}$. Eventually, we end with 12, 22, and 20 entries, respectively. They are displayed in Fig.~\ref{fig5.3}.

An overall anti-correlation between the $\log E_{\rm peak}^{\rm rest}$ and $\beta$ values can be seen. It is strongest in case of the PLC fits, $r=-0.63$ (95\% CI: $(-0.83,-0.28)$)---even stronger than in \citet{dichiara16}. It is weaker for the pure PL case ($r=-0.41$; 95\% CI: $(-0.80,0.21)$---consistent with a lack of correlation), and similar when the set $\beta_{PL} \cup \beta_{PLC}$ is considered ($r=-0.38$, 95\% CI: $(-0.67,-0.05)$). The weakest correlation ($r=-0.21$, 95\% CI: $(-0.60,0.26)$---consistent with a lack of correlation) is attained for $\beta_{2}$, and for the whole set of $\beta$ it is a moderate $r=-0.26$ (95\% CI: $(-0.49,0.01)$---barely consistent with a lack of correlation). We note, however, that (i) our sample is $\sim$2.5 times smaller than that of \citet{dichiara16}, and (ii) we did not impose any conditions on the signal-to-noise ratio, contrary to \citet{dichiara16}. Therefore, in our sample of 54 GRBs with $E_{\rm peak}^{\rm rest}$ values there are GRBs significantly contaminated by the Poisson noise component as well, which likely affects the $\log E_{\rm peak}^{\rm rest}-\beta$ relation. Since the 95\% CIs for $r$ contain (at least marginally) the value describing the sample of \citet{dichiara16}, we do not reject the existence of such correlation (although some subsamples of our $\beta$ allow a lack of correlation, too); however, a bigger sample is definitely required to constrain the relation further, which is outside the scope of this paper.

\subsection{The $L_{\rm iso} - f_0$ relation }
\label{sect3.6}

The peak isotropic luminosity is computed as \citep{schaefer07}
\begin{equation}
L_{\rm iso} = 4\pi d_L^2(z) \mathcal{P} \frac{ \int_{E_1/(1+z)}^{E_2/(1+z)} EN(E)dE }{\int_{E_{\rm min}}^{E_{\rm max}} EN(E)dE},
\label{eq14}
\end{equation}
where $d_L(z)$ is the luminosity distance to a source at redshift $z$, calculated using the latest cosmological parameters within a flat $\Lambda$CDM model \citep{Planck2018}: $H_0=67.4\,{\rm km}\,{\rm s}^{-1}\,{\rm Mpc}^{-1}$, $\Omega_m=0.315$, $\Omega_\Lambda=0.685$; $\mathcal{P}$ is the energy flux (in units of ${\rm erg}\,{\rm cm}^{-2}\,{\rm s}^{-1}$) over the time range of the peak flux of the GRB\footnote{The \texttt{pflx\_xxxx\_ergflux} entries from the Fermi/GBM catalog were employed, where \texttt{xxxx} stands for \textsc{Band}, \textsc{Comp}, \textsc{PLaw}, or \textsc{Sbpl}. }; and $N(E)$ is the spectral model over the time range of the peak flux (expressed in units of $\rm{ph}\,\rm{cm}^{-2}\,\rm{s}^{-1}\,\rm{keV}^{-1}$): \textsc{Plaw}, \textsc{Band}, \textsc{Comp}, or \textsc{Sbpl}, chosen for each GRB according to the \texttt{pflx\_best\_fitting\_model} entry from the Fermi/GBM catalog, and with parameters from therein \citep{kaneko06,gruber}. The integration limits are set using $\{E_1,E_2\}=\{1,10^4\}\,{\rm keV}$, and $\{E_{\rm min},E_{\rm max}\}=\{10,10^3\}\,{\rm keV}$ is the observing bandwidth of Fermi/GBM (cf. Sect.~\ref{sect3.5}). The uncertainties of $\log L_{\rm iso}$ are obtained by bootstrapping the parameters of $N(E)$ \citep{ukwatta10}.

\begin{figure}
\centering
\includegraphics[width=\columnwidth]{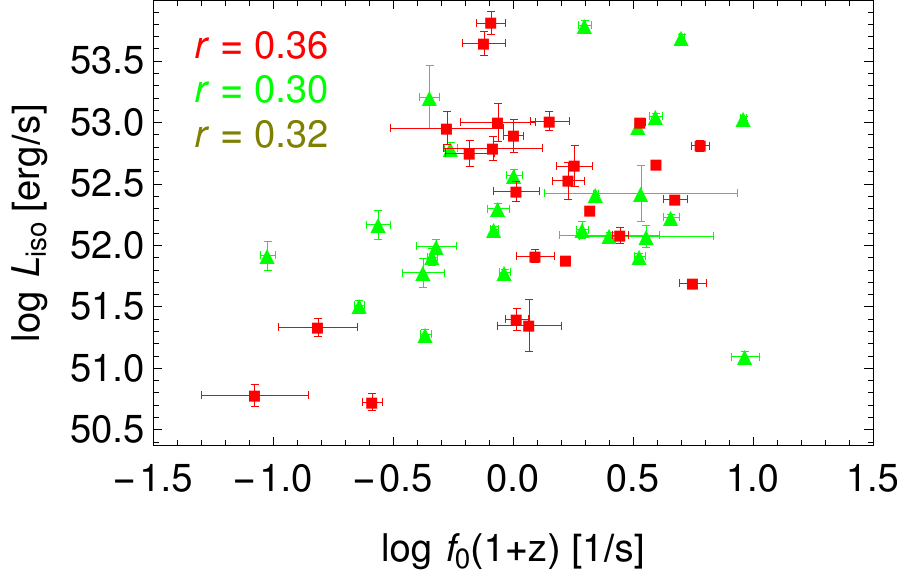}
\caption{ The relation between $L_{\rm iso}$ and the redshift-corrected frequency $(1+z)f_0$. Red squares denote the 25 GRBs whose PSDs are best fitted with a PLC; green triangles correspond to the remaining 26 SBPL fits. The corresponding correlation coefficients $r$ are indicated with respective colors.  }
\label{fig5.4}
\end{figure}

We obtain 81 GRBs with $L_{\rm iso}$ estimates. Hereinafter we employ 25 GRBs that have PSDs best fitted by a PLC, and 26 with an SBPL form. We compare our results (Fig.~\ref{fig5.4}) with those of \citet{ukwatta11} regarding the relation between redshift-corrected characteristic frequency, $(1+z)f_0$, and $L_{\rm iso}$. Our sample yields $r=0.36$ (95\% CI: $(-0.04,0.66)$) and $r=0.32$ (95\% CI: $(0.05,0.55)$) for the PLC and all 51 cases, respectively, while \citet{ukwatta11} obtained $r=0.77$ (95\% CI: $(0.64,0.86)$) with a sample of 58 GRBs.  Our correlation is marginally significant, although weaker and slightly inconsistent with that of \citet{ukwatta11}.

The discrepancy lies in (i) different models employed: \citet{ukwatta11} fitted the PSDs with a piecewise linear function, with a nonzero slope for $f<f_0$, and a constant level for $f>f_0$, while Eq.~(\ref{eq5}) describes a smooth transition. Therefore, our $f_0$ has a slightly different meaning than the $f_0$ of \citet{ukwatta11}. Moreover,  (ii) we did not impose any constraints on the initial sample, as we aimed to analyze the whole Swift catalog, so that more noisy GRBs might be adding variance to the $L_{\rm iso} - f_0$ relation. Note that our sample is about he same size (51 vs. 58) as that of \citet{ukwatta11}.

\section{Discussion}
\label{sect4}

\subsection{PSDs}

The PSDs of Swift GRBs examined herein come in three shapes: a pure PL, a PL with Poisson noise, and an SBPL. The PL case includes flat PSDs, i.e. white noise, which are quite abundant in our sample. They are characteristic of GRBs which are dim, i.e. have a low signal-to-noise ratio, and hence are dominated by the Poisson statistics. The PSDs which are colored noise have indices $\beta\lesssim 2$, so are generally flatter than red noise. When the Poisson noise component becomes significant, the $\beta$ indices in the PLC case rise as well, falling in the range $1\lesssim\beta\lesssim 3$. There is also a non-negligible fraction of steeper PSDs, with $4\lesssim\beta\lesssim 6$, and a few instances of $\beta>6$. The latter basically implies no variability on the corresponding time scales, since it means that for every decade in frequency there is a $>6$ orders of magnitude change in the power. Such steep PLC models ought to be considered artifacts, unless proven otherwise, since we observed they occur when the binned PSD exhibits just one or two points at low frequencies (reflecting the length of the LC), greatly above the Poisson noise level, and hence the fitting results become severely biased. The mode of the $\beta$ distribution in the PLC model is at 1.8, somewhat close to the value $\nicefrac{5}{3}$ expected in the turbulence model. Finally, the SBPL model has its low-frequency index $\beta_1$ gathered around zero, and not exceeding 2, while the high-frequency index $\beta_2$ mostly falls in the range $2\lesssim\beta_2\lesssim 6$. Such steepness is more reliable than in the PLC case, since it occurs at time scales located between the region of low-frequency PL part (well above the Poisson noise level, so detected confidently), and the high-frequency region of Poisson noise dominance. This shows that there is a characteristic time scale, locating the break $T_{\rm break}$between the two PL parts with $\beta_1$ and $\beta_2$, on the order of $1-100$~s, and hence implies there are either two dominant processes working in the progenitors at the emission sight, or---when $\beta_2$ is very steep---the variability at the intermediate time scales is essentially wiped out. The latter can be indicative of a sharp cut-off corresponding to, e.g., the inner edge of the accretion disk.

\subsection{QPOs}

Another feature that was uncovered in some number of GRBs are QPOs---either with an approximately constant leading period, or in the form of up- or down-chirps. As noted in Sect.~\ref{sect1}, there have not been many reports on QPOs in GRBs, hence the instances gathered in Table~\ref{tbl2} are remarkably numerous. The already proposed generation mechanisms (MRI \citep{masada07}; precessing magnetic field \citep{ziaeepour11}) can be complemented with some models employed for active galactic nuclei (AGNs), since both GRBs and AGNs often exhibit striking similarities \citep{wang14,wu16,deng16}. In the simplest scenario, association of the break time scale with the viscous time scale of an accretion disk, coupled with the Keplerian motion on a circular orbit around the newly forming BH \citep{mohan14,zywucka20} can explain the PSD breaks in GRBs as well. Since a relativistic two-body problem (in both Schwarzschild and Kerr metrics) allows for an inspiral (which is impossible in the Newtonian framework), occuring in a finite time \citep{levin08}, a fragmented accretion disk could result in QPOs, lasting several cycles, and possibly chirping signals as well. The orbital period at the innermost stable circular orbit (ISCO) is $T_{\rm ISCO}=12\pi\sqrt{6}GM_{\bullet}/c^3\approx 4.5\cdot 10^{-4}M_{\bullet}/M_{\odot}\,{\rm [s]}$ \citep{hartle}, giving for a typical stellar-mass BH with $M_{\bullet}=10M_{\odot}$ a period of $T_{\rm ISCO}=4.5\,{\rm ms}$---an order of magnitude smaller than the employed binning (64~ms), and smaller than the detected QPO time scales.\footnote{For a Kerr BH with dimensionless spin $a>0$, $T_{\rm ISCO}$ is even smaller \citep{bardeen72}, up to a factor of $3\sqrt{6}$ for a maximally rotating BH.}\textsuperscript{,}\footnote{However, a period on the order of miliseconds is comparable to the QPO discovered in a short BATSE GRB \citep{zhilyaev09}, hence an inspiral inside the ISCO might, at least in some cases, lead to a QPO.} The accretion disk might actually be truncated, with an inner edge at a radius $r=kr_{\rm ISCO}$ ($k\geqslant 1$), in case of which the period $T=k^{\nicefrac{3}{2}}T_{\rm ISCO}$. E.g., $k=20$ changes the 4.5~ms period to 0.4~s---still a few times shorter than the shortest QPO reported in Table~\ref{tbl2}, which would require $k\approx 61$. A plausible range of the cut-off can extend up to $k\sim 100$, giving $T=4.5\,{\rm s}$, consistent with some of the QPOs in Table~\ref{tbl2}. While some QPOs might therefore be indeed due to a truncated disk, it seems unlikely to be a universal explanation. Several more sophisticated orbital models (oscillatory modes in accretion disks, both thin and thick; relativistic precession; tidal disruption (TD) models; warped disk; etc.) were considered in the context of X-ray binaries and microquasars \citep[][and references therein]{torok11,kotrlova20}, and predict the existence of resonant QPOs of a wide range of frequency ratios (cf. fig.~3 in \citealt{kotrlova20}). However, since $\frac{M_{\bullet}}{M_\odot}\frac{f_U}{10^3\,{\rm Hz}}\gtrsim 1$ (cf. fig.~2 in \citealt{kotrlova20}), where $f_U$ is the higher frequency forming the resonant ratio $f_U/f_L>1$, and the ratios in Table~\ref{tbl2} are of the order of unity, the time scales of the QPOs are of the order of 0.01~s. The TD model, in turn, predicts that inhomogeneities with density $\rho$ in the disk will be stretched and disrupted at the Roche limit, and eventually lead to modulation with a period $T_{TD}\sim (G\rho)^{-\nicefrac{1}{2}}$ \citep{cadez08,kostic09,torok11}. Assuming rocky material (planetary/cometary debris) with $\rho=5500\,{\rm kg}\,{\rm m}^{-3}$ (Earth's density), $T_{TD}=1650\,{\rm s}$. Lower $\rho$ gives higher $T_{TD}$. To match the 10~s QPO period, $\rho\sim 10^8\,{\rm kg}\,{\rm m}^{-3}$, which is an unlikely possibility.  

The relativistic motion around Kerr BHs can lead to even more complicated, three dimensional orbits, giving rise to breaks as well as QPOs \citep{rana19,rana20b}. The low-frequency QPOs, with time scales $1-10\,{\rm s}$, in fact arise naturally in this setup, hence appear to be a probable description for the QPOs in GRBs (cf. table~10 in \citealt{rana20b}), and account for resonant QPOs as well. Finally, the Lense-Thirring precession leads to frequencies matching the QPOs when the disk is truncated at $k\gtrsim 30$ (cf. fig.~5 in \citealt{ingram09}), an order of magnitude smaller than in the above orbital models.

A detailed picture was painted with the use of magnetohydrodynamical simulations of a forced perturbance within a magnetized accretion disk \citep{petri05}. This scenario seems plausible, since among the 10 chirps we identified, 8 are up-chirps, i.e. with a decreasing period. The two down-chirps would require a different mechanism, though. An appealing one might be due to a helical jet, which when applied to AGNs, results in down-chirps \citep{mohan15}. On the other hand, accretion flows in which MRI dominates do not exhibit QPOs, since MRI turbulence destroys coherence, hence nullifies QPOs, while in a magnetically choked accretion flow (with an accumulated magnetic flux), there appear QPOs with periods $\sim 70GM_{\bullet}/c^3\approx 0.004\,{\rm s}$ for $M_{\bullet}=10M_\odot$ (see \citealt{mckinney12} for details), i.e. again too low to match the QPOs detected in GRBs. Finally, oscillations of the shock front can give rise to QPOs with Hz and sub-Hz frequencies, i.e. could possibly explain some of the QPOs in GRBs as well \citep[][and references therein]{iyer15,palit19}.

GRB emission is of synchrotron nature \citep{burgess20,ghisellini20}, and comes from the shocks in relativistic jets. We discussed possibilities of generating QPOs of appropriate periods in the surrounding disk (except for the shock front oscillations), assuming they transfer with a one-to-one correspondence to the jet via disk-jet coupling. This might not be entirely true, and/or the observed QPOs might as well arise due to a combination of more than one effect.

\subsection{Persistence}

A conservative methodology used to estimate the Hurst exponents revealed that 93\% of GRBs are characterized by $H>0.5$, meaning they exhibit long-term memory, or persistence. Recall that the value of $H$ can be attributed to both stationary (e.g., colored noise with $\beta<1$ or fractional Gaussian noise) and nonstationary (e.g., PL with $\beta>1$ or fractional Brownian motion), and hence the notion of smoothness it quantifies is broader that (non)stationarity. We therefore showed that the autocorrelations in the GRBs' variability persist throughout the LCs, hence---so to speak---the random component embedded in the $\gamma$-ray signal is structured on a fundamental level.

\subsection{$\mathcal{A}-\mathcal{T}$ plane}

We attempted also to classify the GRB prompt LCs in the recently developed $\mathcal{A}-\mathcal{T}$ plane. We considered here only long GRBs, with $T_{100}>3.2\,{\rm sec}$, and since the dichotomy between short and long GRBs is well established, we expected to verify with yet another approach the existence of the presumed third, intermediate class of GRBs. We indeed observed hints of clustering in two regions of the $\mathcal{A}-\mathcal{T}$ plane. However, one of the groups tends to gather around the point $(1,\nicefrac{2}{3})$, i.e. the location of white noise processes. When examined in dependence on the $\beta$ indices of the PSDs, it turned out that area is occupied by PLC and SBPL models with very high values of $\beta$---i.e., those GRBs predominantly exhibitng white noise PSDs. As noted above, such cases are either due to the objects being dim, spurious fits, or a nontrivial coexistence of two white noise processes represented by components with different powers. We therefore conclude there are no unambiguous signs of a sub-classification of long GRBs' LCs. While this is not a definite proof of the non-existence of the third class on its own, it is consistent with other works tackling this issue more directly \citep{tarnopolski16a,tarnopolski19a,tarnopolski19c,jespersen20}.

\subsection{Correlations}

Finally, we critically revisited the $E_{\rm peak}^{\rm rest}-\beta$ and $L_{\rm iso} - f_0$ relations, confirming their existence in the whole Swift catalog. Comparing with the corresponding results of \citet{dichiara16} and \citet{ukwatta11}, we obtained slightly stronger and significantly weaker, respectively, correlations in the appropriate relations. In case of the $L_{\rm iso} - f_0$ relation, the overall positive correlation might in fact be a simple luminosity effect: the brighter the source, the higher its signal-to-noise ratio, hence the lower the contamination of the signal with Poisson noise. This then implies that the location of the critical frequency, $f_0$, is shifted to higher frequencies, making the white noise component less significant. This is backed up also by a very strong correlation ($r=0.98$) between the Poisson noise levels obtained from fits and directly from the uncertainties of the LC measurements.

The $E_{\rm peak}^{\rm rest}-\beta$ relation, in turn, connects the spectral energetic properties of a GRB with a characteristic of an LC. An anticorrelation between the two might be a sign of a common, physical parameter governing the values of $E_{\rm peak}^{\rm rest}$ and $\beta$, e.g. the $\Gamma$ factor.

\section{Summary}
\label{sect5}

\begin{enumerate}
\item The PSDs with PL and PLC shapes are broadly consistent with the $\nicefrac{5}{3}$ Kolmogorov law expected in a fully developed turbulence. Several cases of an SBPL model were also obtained, with break time scales on the order of 1--100~s.
\item We reported on QPOs detected in the wavelet scalograms of 34 GRBs: 10 chirping signals (8 up-chirps and 2 down-chirps), and 24 QPOs with constant leading periods (13 with a single QPO, 8 with two coexisting QPOs, and 3 triple QPOs). In particular, we confirmed a persistent QPO with an $\sim$8~s period in GRB 090709A. We identify non-planar orbits around Kerr BHs, the Lense-Thirring effect, and shock oscillations as plausible mechanisms for the QPO generation.
\item 93\% of GRBs are characterized by $H>0.5$, i.e. they express long-term memory in the prompt LCs, not connected trivially with the PSD features.
\item The $\mathcal{A}-\mathcal{T}$ plane did not reveal any meaningful separation of long GRBs into subclasses.
\item The $E_{\rm peak}^{\rm rest}-\beta$ and $L_{\rm iso} - f_0$ relations were confirmed, though the latter seems to be a straightforward result of the luminosity effect.
\end{enumerate}

\acknowledgments
We thank Agnieszka Janiuk for useful comments on the draft, and the anonymous reviewer for helpful suggestions. M.T. acknowledges support by the Polish National Science Center (NSC) through the OPUS grant 2017/25/B/ST9/01208. V.M. was supported by the NSC grant 2016/22/E/ST9/00061.

\software{\textsc{Mathematica} \citep[v10.4; ][]{Mathematica}, \textsc{SciPy} \citep[v1.1.0; ][]{scipy}, \textsc{wavepal} \citep{lenoir18a,lenoir18b}}.

\bibliography{bibliography}{}

\end{document}